\documentclass[english
]{aa}
\usepackage{epsf,amsfonts,amssymb,graphicx,fancyheadings,caption}
\usepackage{natbib}
\usepackage{babel}
\usepackage{txfonts}

\bibpunct{(}{)}{;}{a}{}{,}

\begin{document}

\title{R Coronae Borealis stars in the Galactic Bulge discovered by EROS-2
\thanks{Based on observations made with the CNRS/INSU MARLY telescope at the European
Southern Observatory, La Silla, Chile.}
}

\author{
P.~Tisserand\inst{1,2},
J.B.~Marquette\inst{3},
P.R.~Wood\inst{1},
\'{E}.~Lesquoy\inst{2,3},
J.P.~Beaulieu\inst{3},
A.~Milsztajn\inst{2}\thanks{deceased},
C.~Hamadache\inst{2},
C.~Afonso\inst{2}\thanks{Now at Max-Planck-Institut f\"ur Astronomie, Koenigstuhl 17, D-69117 Heidelberg,
 Germany},
J.N.~Albert\inst{4},
J.~Andersen\inst{5},
R.~Ansari\inst{4},
\'{E}.~Aubourg\inst{2},
P.~Bareyre\inst{2},
X.~Charlot\inst{2},
C.~Coutures\inst{2,3},
R.~Ferlet\inst{3},
P.~Fouqu\'{e}\inst{6},
J.F.~Glicenstein\inst{2},
B.~Goldman\inst{2}\thanks{Now at Max-Planck-Institut f\"ur Astronomie, Koenigstuhl 17, D-69117
 Heidelberg, Germany},
A.~Gould\inst{8},
M.~Gros\inst{2},
J.~Haissinski\inst{4},
J.~de Kat\inst{2},
L.~Le Guillou\inst{2}\thanks{Now at LPNHE, IN2P3 CNRS and Universit\'{e}s Paris 6 \& Paris 7, 4 place
 Jussieu, 75252 Paris Cedex 05, France},
C.~Loup\inst{3}\thanks{Now at Observatoire Astronomique de Strasbourg, UMR 7550, 11 rue de
 l'Universit\'e, 67000 Strasbourg, France},
C.~Magneville\inst{2},
\'{E}.~Maurice\inst{9},
A.~Maury\inst{7}\thanks{Now at San Pedro de Atacama Celestial Exploration,
Casilla 21, San Pedro de Atacama, Chile},
M.~Moniez\inst{4},
N.~Palanque-Delabrouille\inst{2},
O.~Perdereau\inst{4},
Y.~Rahal\inst{4},
J.~Rich\inst{2},
M.~Spiro\inst{2},
A.~Vidal-Madjar\inst{3},
and S.~Zylberajch\inst{2}
}

\institute{
Research School of Astronomy and Astrophysics, Australian National University, Cotter Rd, Weston Creek
 ACT 2611, Australia \and
CEA, DSM, DAPNIA, Centre d'\'{E}tudes de Saclay, 91191 Gif-sur-Yvette Cedex, France \and
Institut d'Astrophysique de Paris, UMR7095 CNRS, Universite Pierre \& Marie Curie, 98 bis boulevard
 Arago, 75014 Paris, France \and
Laboratoire de l'Acc\'{e}l\'{e}rateur Lin\'{e}aire, IN2P3 CNRS, Universit\'{e} de Paris-Sud, 91405 Orsay
 Cedex, France \and
The Niels Bohr Institute, Astronomy Group, Juliane Maries Vej 30, 2100 Copenhagen, Denmark \and
Observatoire Midi-Pyr\'en\'ees, UMR 5572, 14 avenue Edouard Belin, 31400 Toulouse, France \and
European Southern Observatory, Casilla 19001, Santiago 19, Chile \and
Department of Astronomy, Ohio State University, Columbus, OH 43210, U.S.A. \and
Observatoire de Marseille, 2 place Le Verrier, 13248 Marseille Cedex 04, France
}

\offprints{Patrick Tisserand; \email{tisserand@mso.anu.edu.au}}

\date{Accepted for publication in A\&A}

\abstract {Rare types of variable star may give unique insight into short-lived stages of stellar evolution. The systematic monitoring of millions of stars and advanced light curve analysis techniques of microlensing surveys make them ideal for discovering also such rare variable stars. One example is the R Coronae Borealis (RCB) stars, a rare type of evolved carbon-rich supergiant.}
{We have conducted a systematic search of the EROS-2 database for the Galactic catalogue Bulge and spiral arms to find Galactic RCB stars.}
{The light curves of $\sim$100 million stars, monitored for 6.7 years (from July 1996 to February 2003), have been analysed to search for the main signature of RCB stars, large and rapid drops in luminosity. Follow-up spectroscopy has been used to confirm the photometric candidates.}
{We have discovered 14 new RCB stars, all in the direction of the Galactic Bulge, bringing the total number of confirmed Galactic RCB stars to about 51.}
{After reddening correction, the colours and absolute magnitudes of at least 9 of the stars are similar to those of Magellanic RCB stars. This suggests that these stars are in fact located in the Galactic Bulge, making them the first RCB stars discovered in the Bulge. The localisation of the 5 remaining RCBs is more uncertain: 4 are either located behind the Bulge at an estimated maximum distance of 14 kpc or have an unusual thick circumstellar shell; the other is a DY Per RCB which may be located in the Bulge, even if it is fainter than the known Magellanic DY Per. From the small scale height found using the 9 new Bulge RCBs, $61<h^{RCB}_{Bulge}<246$ pc (95\% C.L.), we conclude that the RCB stars follow a disk-like distribution inside the Bulge.}


\authorrunning{Tisserand, P. et al.}
\titlerunning{14 new Galactic R Coronae Borealis stars from EROS-2}

\maketitle

\section{Introduction \label{sec_intro}}

The R Coronae Borealis (RCB) stars constitute a rare type of hydrogen deficient, carbon-rich supergiant star. Only 38 examples are currently known in the Galaxy \citep{1996PASP..108..225C, 2005AJ....130.2293Z}, 17 in the Large Magellanic Cloud \citep{2001ApJ...554..298A} and 6 in the Small Magellanic Cloud \citep{2004A&A...424..245T, 2005ApJ...631L.147K}. This rarity presumably indicates a brief phase of stellar evolution. A detailed review of their characteristics has been written by \citet{1996PASP..108..225C}. RCBs exhibit spectacular, unpredictable, rapid declines in brightness (up to 9 magnitudes in optical wavelengths over a few days or weeks) thought to be due to the photosphere being obscured by newly formed dust clouds along the line of sight. As the dust clouds disperse, the original brightness gradually recovers in months. Several of these clouds have been directly observed in various directions around the RCB star RY Sgr, see \citet{2004A&A...428L..13D, 2007A&A...466L...1L}.

Up to now, there has been no reliable distance estimate to any Galactic RCB star, so their absolute magnitudes is only known from the RCBs discovered in the Magellanic Clouds (MC). A direct relation between effective temperature and absolute magnitude was observed by \citet{2001ApJ...554..298A}. Hipparcos gave lower limits to the distances to several of the brightest Galactic RCB stars \citep{1997PASP..109.1089T, 1998PASA...15..179C}. Such limits imply that the RCB stars studied must be brighter than $M_V = -3$. 

The true spatial distribution of RCB stars is still uncertain. This is due to the small number of known RCB stars, which is also biased toward higher Galactic latitudes as reddened stars in the Galactic plane can be missed in magnitude limited surveys \citep{1990MNRAS.247...91L, 1990ASPC...11..566L}. Different views are found in the literature. First, \citet{1985ApJS...58..661I} reports a scale height of $\sim400$ pc for known RCB stars assuming $M_{Bol} = -5$. So RCB stars may be part of an old disk-like distribution. This idea is also supported by \citet{2005AJ....130.2293Z}, who mention a possible thick-disk distribution. However, \citet{1998PASA...15..179C} noted that the Hipparcos velocity dispersion of RCB stars is quite similar to those of other cool hydrogen-deficient carbon stars and extreme Helium (eHe) stars, suggesting that RCB stars might have a Bulge-like distribution.

Two major evolutionary scenarios are suggested to explain their origin : the Double Degenerate scenario (DD) and the final Helium Shell Flash (FF) scenario \citep{1996ApJ...456..750I, 1990ASPC...11..549R}. The DD model involves the merger of a CO- and a He- white dwarf and was recently strongly supported by the observations of $^{18}$O over-abundance in seven H-deficient carbon and RCB stars, that is not expected in the FF model \citep{2007ApJ...662.1220C}. The birthrates of RCB stars would then help us to better understand the rates of mergers for objects that could be Supernovae type Ia progenitors \citep{1984ApJ...277..355W,2005ApJ...629..915B}. The FF model involves the expansion of a star, on the verge of becoming a white dwarf, to a supergiant size. This outburst phenomena has already been observed in three stars (Sakurai's object, V605 Aql and FG Sge), which transformed themselves into cool giant stars with spectral properties similar to those of RCB stars \citep{1997AJ....114.2679C, 1999A&A...343..507A,1998ApJS..114..133G}. However, \citet{2006ApJ...646L..69C} note that differences between FF star and RCB star light curve variations and abundance patterns (such as $^{12}$C$/^{13}$C ratio) indicate that the former are unlikely to be the evolutionary precursors of the majority of the latter.

One way to make progress in the comprehension of these stars is to enlarge the small sample of known objects. Microlensing surveys (OGLE, MACHO, MOA, EROS and Supermacho) are ideal for increasing the sample because they monitor millions of stars and thus could observe the characteristic RCB signature of rapid fading. The EROS-2 (Exp\'erience de Recherche d'Objets Sombres) experiment is currently the survey that has monitored the largest sky area ($\sim106$ deg$^2$ in our Galaxy) during a period of $\sim$6.7 years. EROS can thus be used to significantly increase the number of known Galactic RCBs.

This article reports a sample of 14 new RCB stars. The photometric and spectroscopic data used are presented in Section~\ref{sec_obs}. A discussion of the previously known RCB stars in and near the area monitored by EROS-2 is given in Section~\ref{knownRCB}. Following the discovery of the first RCB stars in the SMC \citep{2004A&A...424..245T}, we tried to increase our detection efficiency to prevent a selection effect due to reddening. The modified detection techniques, and the spectroscopic observations, are described in Section~\ref{mining.sect}. Finally, the general characteristics of the new RCBs discovered are discussed in Sections~\ref{sec_NewRCB} and \ref{summary}.

\begin{table*} 
\caption{Known Galactic RCB stars located in the EROS-2 fields
\label{tab.KnownRCB.GenInfo}}
\medskip
\centering
\begin{tabular}{llllc}
\hline
\hline
Star name & Coordinates ($J_{2000}$) & True MACHO Id & Situation on EROS-2 & Large variations\\
 & & & reference image & seen by EROS-2?\\
\hline
\object{V1783 Sgr}   &  18:04:49.74 -32:43:13.6  & & saturated, cg0842m & no\\ 
\object{V739 Sgr}   &  18:13:10.54 -30:16:14.7  & \object{MACHO-123.24377.7} & saturated, cg6245k & yes\\
\object{V3795 Sgr}  &  18:13:23.58 -25:46:40.8  & \object{MACHO-161.24445.6} & saturated, cg6273l & no\\
\object{VZ Sgr}      &  18:15:08.58 -29:42:29.4  & \object{MACHO-117.25166.4693} & saturated, cg0290n & yes \\
\object{MACHO-135.27132.51} &  18:19:33.87 -28:35:57.8  & & too faint, cg0353k & yes\\
\hline
\end{tabular}
\end{table*}

\section{Observational data}
\label{sec_obs}
The EROS-2 project used the 1-meter MARLY telescope at ESO La Silla Observatory, Chile. The primary purpose of the project was to search for microlensing events \citep{1986ApJ...304....1P} due to baryonic dark matter in the halo \citep{2007A&A...469..387T} or to ordinary stars in the Galactic plane \citep{2006A&A...454..185H,2001A&A...373..126D}. The observations were performed between July 1996 and February 2003 with two wide field cameras (0.69\degr in right ascension $\times$ 1.38\degr in declination, thus a sky area of $\sim$0.95 deg$^2$) behind a dichroic cube splitting the light beam into two broad passbands. The so-called ``blue'' channel (420-720 nm, hereafter magnitudes $B_\mathrm{E}$) overlapped the $V$ and $R$ standard bands while the ``red'' one (620-920 nm, hereafter magnitudes $R_\mathrm{E}$) roughly matched the mean wavelength of the Cousins $I$ band. Each camera was a mosaic of eight 2048 $\times$ 2048 CCDs with a pixel size of 0.6\arcsec ~on the sky.

The photometric calibration was obtained directly for 20\% of our fields by matching our star catalogues with those of the OGLE-II collaboration \citep{2002AcA....52..217U}. To a precision of 0.1 mag, we found the following transformations with the standard V Johnson and I Cousins broadband:
\begin{equation}
$$ R_E = I, \hspace{3 mm} B_E = V - 0.4(V - I)$$.
\label{eq.photcalib}
\end{equation}
The calibration is sufficiently uniform that it can be extended with confidence to the remaining fields. We also note that the OGLE calibration used a colour range of observed standard stars (Landolt system, \citet{1992AJ....104..372L}) limited to $V-I < 2$ mag. For redder stars, they stressed that their transformations are extrapolations with a systematic error that may reach 0.25 mag at the very red edge ($V-I>4 $ mag). This is of particular interest for our study.

Five distinct sky areas have been monitored in our galaxy. The main sky area corresponds to 83 fields centred on the Galactic Centre (60 fields in the south Galactic side (b$<$0) and 23 in the north one). These bulge fields are designated ``cgxxxqq'' where ``xxx'' is the number of the field and ``qq'' is the quadrant of a CCD. The remaining four zones are distributed in crowded regions along the spiral arms. They cover a surface of $\sim27.5$ deg$^2$ with 29 fields and are listed individually in Table~\ref{tab.fields}. The photometry of individual images and the reconstruction of the light curves were processed using the Peida package which has been specifically developed for the EROS experiment \citep{1996VA.....40..519A}. A detailed discussion about the photometric accuracy can be found in \citet{2006A&A...454..185H} for the Bulge fields and in \citet{Rahal2003th} for the spiral arms ones. The photometric precision for Bulge clump-giant stars is better than 2\% for $R_\mathrm{E} < 17$ and about 4\% for $R_\mathrm{E}\sim18$.

\begin{table} 
\caption{Galactic sky areas monitored by EROS-2.
\label{tab.fields}}
\medskip
\centering
\begin{tabular}{lllll}
\hline
\hline
Sky area  & EROS2 & Number & \\
& identifier & of fields & $\left\langle l\right\rangle$ & $\left\langle b\right\rangle$ \\
\hline
Galactic Centre & cgxxx & 83 & 1.5 & -3.0 \\
$\gamma$ Scuti & gsxxx & 5 & 18.71 & 2.66 \\
$\beta$ Scuti & bsxxx & 6 & 27.75 & 2.64 \\
$\gamma$ Normae & gnxxx & 12 & 331.38 & 2.85 \\
$\theta$ Muscae & tmxxx & 6 & 306.85 & 1.78 \\
\hline
\end{tabular}
\end{table}

The template images are formed with 15 good seeing co-added images. They are used to detect sources and then form the initial object catalogue. With an exposure time of 120 seconds (except $\theta$ Muscae, 180s), our detection limit reach about the 21st magnitude while stars brighter than the $\sim$11th magnitude are saturated. An average of one point every four nights was taken for each field during the Galactic Bulge observing season though the sampling frequency occasionally reached more than one point per night at mid-season. Stars were imaged in average on 800, 550 and 350 epochs for the cg, tm-gn and bs-gs fields respectively. Some low priority cg fields (18 out of the 83), less frequently observed (from $\sim$80 to $\sim$200 epochs), have also been considered in this analysis.


Spectroscopy of RCB candidates was performed with the Dual-Beam Spectrograph (DBS) \citep{1988PASP..100..626R} attached on the  ANU\footnote{Australian National University} 2.3m diameter telescope based at Siding Spring Observatory. The DBS is a general purpose optical spectrograph, permanently mounted at the Nasmyth A focus. The visible wave-band is split by a dichroic at around 600 nm and feeds two essentially similar spectrographs, with red and blue optimised detectors respectively. The full slit length is 6.7 arcmin. Observations in the redder part (530 to 1000 nm) are presented, with a 2-pixel resolution of 2 \AA{} or 7.5 \AA{}.

\section{Previously known RCB stars}
\label{knownRCB}

38 spectroscopically confirmed RCB stars are known in our galaxy : 31 RCBs\footnote{We do not include V1773 Oph, GM Ser and V1405 Cyg as RCB stars as they have never been spectroscopically confirmed or shown photometric variations larger than 2.5 mag, see \citet{1997Obs...117..205K} and \citet{1990Ap&SS.172..263M}.} were listed by \citet{1996PASP..108..225C} and 7 others were recently discovered \citep{2002PASP..114..846C,2003PASP..115.1301H,2005AJ....130.2293Z}. Fourteen of them lie in or near the area monitored by EROS-2.


Of the fourteen, eight are outside the EROS-2 fields : the newly active warm RCB V 2552 Ophiuchi \citep{2003PASP..115.1301H}, V517 Oph, GU Sgr, V348 Sgr, WX Cra and 3 RCBs recently discovered by  \citet{2005AJ....130.2293Z} in the MACHO database (MACHO-308.38099.66, MACHO-301.45783.9 that lies between 2 CCDs in the spiral arm field gs202, and MACHO-401.48170.2237 between Bulge fields cg0084l and cg0051m).

The 6 others, which lie in the EROS-2 fields, are individually discussed here. Three of them (V739 Sgr, V3795 Sgr and VZ Sgr) were already listed as RCB stars by \citet{2005AJ....130.2293Z}, but with wrong coordinates and thus also wrong MACHO identifiers. The correct values are listed in Table~\ref{tab.KnownRCB.GenInfo}. We also indicate in this Table that three of those six RCB stars (V739 Sgr, VZ Sgr and MACHO-135.27132.51) showed large variations during the EROS-2 observation, but were not catalogued because they were either saturated or too faint on our reference images. For these stars, we then re-processed all our available EROS-2 data with our PSF photometry pipeline and new reference images to reconstruct their light curve. These re-processed light curves are presented in Figure~\ref{lc_KnownRCB}.

\begin{figure} 
\centering
\resizebox{\hsize}{!}{\includegraphics{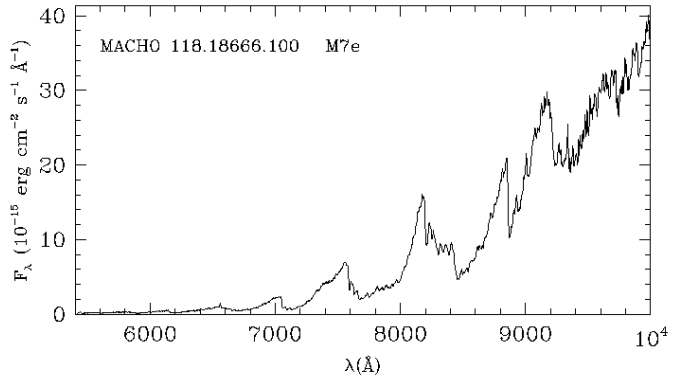}}
\caption{Spectrum of MACHO-118.18666.100.}
\label{MACHO118_spectrum}
\end{figure}

\begin{itemize} 
 \item V739 Sgr: This star was reported as a possible RCB star by \citet{1991Obs...111..244L} due to its carbon spectrum and observed long-term photometric variability of large amplitude ($\sim4.5$ mag). It has also been listed in the cool carbon star catalogue by \citet{1989PW&SO...3...53S}. A smooth symmetric variation of 2.5 mag that lasts $\sim120$ days is observed in its MACHO light curve.

 V739 Sgr was saturated on the EROS-2 reference images (subfield cg6245k) and has thus not been catalogued. Its re-processed light curve shows 2 drops of $\sim8$ mag (Fig.~\ref{lc_KnownRCB}). These declines are characteristic of an RCB light curve.

 \item V3795 Sgr: This RCB star remained saturated in all EROS-2 epochs (subfield cg6273l). Its AAVSO light curve show 2 declines, one before the EROS-2 observations ($HJD\simeq-1500$) and one after ($HJD\simeq2700$), both larger than 3 mag.

The recovery phase of the first decline is also visible in the MACHO light curve, the second in the ASAS-3\footnote{ASAS: All Sky Automated Survey \citep{1997AcA....47..467P}, URL:  http://www.astrouw.edu.pl/$\sim$gp/asas/asas.html} V band light curve. We note that V3795 Sgr was saturated most of time in the MACHO light curve, except for the portion below maximum light.

 \item VZ Sgr: The neighbourhood of this RCB star is very crowded. Its MACHO light curve shows a decline of 5 mag in about 20 days around JD-2449180. The MACHO curve consists of only 11 measurements, VZ Sgr being located in a low-sampling MACHO field.

VZ Sgr is also located in a low-sampling EROS-2 field (cg0290n) and was saturated on our reference images. Its re-processed light curve shows variations of $\sim6$ mag (Fig.~\ref{lc_KnownRCB}). The MACHO and EROS variations can be better understood by comparison to the well sampled AAVSO light curve showing multiple declines since $\sim2446000$.

 \item V1783 Sgr: This star was reported as a possible RCB star by \citet{1991Obs...111..244L} due to its H-poor carbon spectrum. It remained saturated in all EROS-2 epochs (located on subfield cg0842m). A slow decline (dV/dt$\sim0.009$ mag.$day^{-1}$) occurred only a few days after the end of the EROS-2 observations (see its ASAS and AAVSO light curve).

 \item MACHO-135.27132.51: This RCB star, located in a low-sampling EROS-2 field (cg0353k), was too faint on our reference images and was thus also not catalogued. However, its re-processed light curve shows multiple variations of $\sim8$ mag and one fast decline (Fig.~\ref{lc_KnownRCB}).

\item MACHO-118.18666.100: This star was properly catalogued and monitored in the EROS-2 database (named EROS2-cg6075k26836). Its light curve is presented in Figure~\ref{lc_KnownRCB}. No drop in luminosity is observed, only some periodic oscillation features. Its classification as an RCB star being debatable, we took a spectrum during the night JD-2454348. The spectrum, presented in Figure~\ref{MACHO118_spectrum},  is typical of an M giant star, type M7e, with TiO features observable. Our spectrum is so different to the one presented by \citet[Fig.3]{2005AJ....130.2293Z} that, obviously, two different stars were observed. We are confident with our observation. We do not then support the classification of MACHO-118.18666.100 as an RCB star. Furthermore, we suggest that the $\sim1.2$ mag variation in $\sim150$ days observed in the MACHO light curve \citep[Fig.1e]{2005AJ....130.2293Z} and the smaller one observed in the EROS2 light curve (see Fig.~\ref{lc_KnownRCB}, JD~2451900) are probably due to a sequence-D variation in a long secondary period \citep[see][]{1999IAUS..191..151W}. A primary pulsation period, characteristic of such stars, is also seen.
\end{itemize}

In summary, six stars previously classified as RCBs are located in the EROS-2 fields; their status is summarized in Table~\ref{tab.KnownRCB.GenInfo}. One of these stars is too faint and four others very bright and saturated in our reference images. Therefore, only one of the six was catalogued and monitored in the EROS-2 database, but its spectrum shows that it is not an RCB star, but an M giant. Overall, we failed to detect any of these known RCB stars. We also note that of those five, only three (MACHO-135.27132.51, VZ Sgr and V739 Sgr) showed detectable luminosity variations. Our light curve analysis would have detected them, had they been catalogued.

In view of this result, the completeness of our search is discussed in detail in Section~\ref{detect_sect}.

\section{Mining the EROS-2 database}
\label{mining.sect}
 The initial object catalogue used in the present analysis is an extended version of the cleaned EROS-2 source star catalogue produced by \citet{2006A&A...454..185H} and \citet{Rahal2003th} for the microlensing analysis. We decided to enlarge our cleaning selection criteria for two reasons. First, over the period covered by the reference images, the RCB stars might be in a high luminosity phase, causing saturation in one band, or in a low luminosity phase, making the star undetectable in one band. Second, the low latitude fields are highly absorbed, making some RCBs undetectable in the blue band. Because of this, we decided to consider as valid all objects detected in at least one of the two filters. This strategy allows us to find RCBs that may have been catalogued in only one filter. Unfortunately, most of the new objects are due to light curves affected by diverse optical or electronic artefacts.

About $\sim$82 and $\sim$25 million different objects are in the extended catalogues of the Galactic Centre and spiral arm fields. They all passed our simple artefact cleaning procedure that requires a minimum average flux of 300 ADU ($R_E\sim20.7$ and $B_E\sim21.4$) and a minimum of 30 measurements per light curve.


Two different strategies, described in the following sections, have been used to find RCB stars in the two catalogues. For both strategies, we search for the outstanding characteristic of RCBs: a rapid drop in luminosity, by $\sim2 - 8$ mag. The first one is based on a series of selection cuts applied to each light curve, both filters being considered separately. The second is a visual inspection of the light curves of all EROS-2 objects catalogued either as a carbon star in the CGCS\footnote{CGCS : The Catalogue of Galactic Cool Carbon Stars, including 6891 entries collected from the literature.} catalogue \citep{2001BaltA..10....1A} and the Bulge carbon stars found by \citet{1991A&AS...88..265A} or with a high infrared excess in the 2MASS database \citep{2006AJ....131.1163S}. The second strategy was designed to verify the results of the first.

\begin{figure} 
\centering
\resizebox{\hsize}{!}{\includegraphics{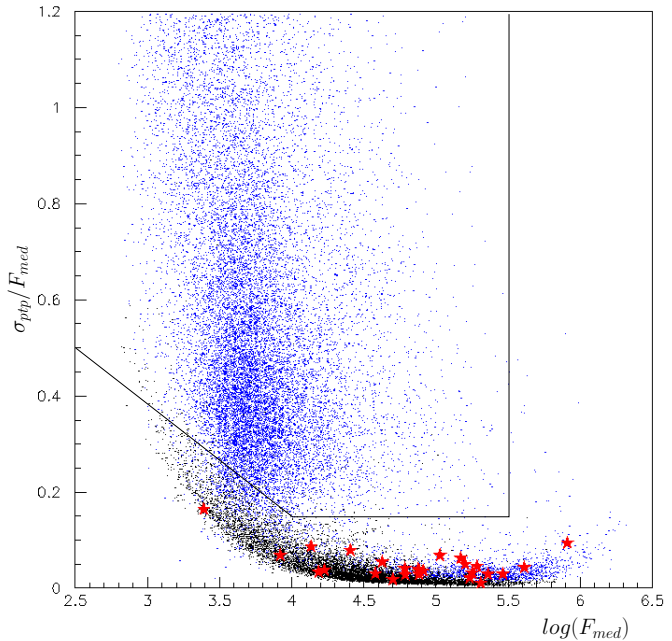}}
\caption{Ratio R ($ = \sigma_{ptp} / F_{med}$) vs. median magnitude $F_{med}$. The blue dots show the 28391 Bulge and 15731 spiral arm objects that pass cuts 1, 2 and 3. The black dots are random stars from one of our sky areas; they show significantly lower point-to-point dispersion at given $F_{med}$. We retain only objects below or to the right of the straight lines. The 28 red stars show the position of the 14 new RCBs in both filters.}
\label{sigmaint_fig}
\end{figure}

\subsection{Light curve analysis}

RCB optical light curves exhibit fast luminosity declines, at irregular intervals, followed by slow recoveries. Multiple consecutive fadings can be observed before the original brightness is recovered. Superimposed on these variations are intrinsic pulsations which produce low-amplitude periodic variations (for a good example, see the light curve of the R Coronae Borealis star prototype, R CrB, monitored since 1843 by the AAVSO\footnote{American Association of Variable Star Observers, URL: http://www.aavso.org/vstar/vsots/0100.shtml}).

Guided by these properties, RCB candidates were selected by requiring that their light curves satisfy 4 quantitative criterion given below. The light curves satisfying these criteria were then visually scanned to establish a list of spectroscopic targets.

The selection criteria were:
\begin{enumerate} 
\item The difference between maximum and minimum flux, $F_{max}$ and $F_{min}$, must exceed 2.5 magnitudes. $F_{max}$ and $F_{min}$ are determined as the averages of the 10 extreme values after eliminating the five highest and lowest flux points.)
\item The maximum flux must be higher than 5000 ADU, equivalent to a magnitude at maximum of about 18 (here, the maximum flux is averaged over 10 consecutive points).
\item The third selection criterion was designed to reject most short-period variable stars. We retain objects whose light curve crosses the median flux $F_{med} = 1/2\times(F_{max} + F_{min})$ by at least 3 consecutive points, less than 12 times.
\item After application of the first three selection criteria, 28391 objects remained in the Bulge fields and 15731 in the spiral arm fields. The light curves of these objects are mainly due to artefacts and show irregular/erratic variations. A fourth criterion was therefore designed to eliminate light curves dominated by artefacts, based on the ratio R between the point-to-point dispersion of the light curve, $\sigma_{ptp}$, and the median flux, $F_{med}$.  Here, $\sigma_{ptp}$ is defined as the dispersion of the flux of individual points from the flux obtained by linear interpolation between the two neighbouring points. Figure~\ref{sigmaint_fig} illustrates the cuts applied in the plane of $\sigma_{ptp}/F_{med}$ vs. $\log F_{med}$ to remove artefact-dominated light curves, based on the locus of a random sample of ``normal'' stars.
\end{enumerate}


The remaining 8524 stars (7181 in the cg field and 1343 in gs, gn, tm and bs, see Table~\ref{tab.fields}) are mainly Mira and other Long-Period Variables. 23 of these stars were selected as RCB candidates by visual inspection, based on the main characteristic of RCB stars, a rapid and non-periodic drop in luminosity. But because our irregular sampling may mislead our interpretation, we decided to act generously. Thus, not all 23 RCB candidates have light curves with a large, rapid drop in luminosity. Half of them have only an infra-red excess ($J-K>1$) and large variations with no clear indication of periodicity, so their classification as RCB stars was considered uncertain. Hence, spectroscopy was obtained for these 23 stars, with results as described in Section~\ref{spect_sect}.


\begin{figure*} 
\sidecaption
\includegraphics[width=9.7cm]{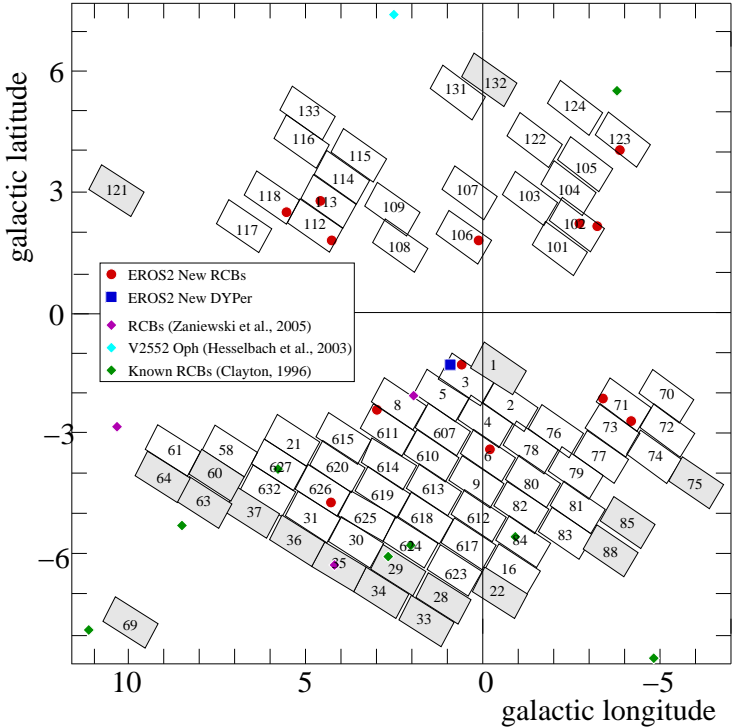}
\caption{Map of the 83 EROS-2 Bulge fields in Galactic coordinates. The 18 less frequently observed fields are coloured grey. The location of the newly confirmed Galactic RCBs stars are indicated along with the four discovered by \citet{2005AJ....130.2293Z}, the newly active warm RCB V 2552 Ophiuchi \citep{2003PASP..115.1301H} and the other previously known RCBs in this region (V1783 Sgr, V739 Sgr, V3795 Sgr, VZ Sgr, V517 Oph, Gu Sgr, V348 Sgr and WX Cra).}
\label{cgfields}
\end{figure*}

\subsection{Detection efficiency}
\label{detect_sect}

The unpredictable nature of RCB light curves, our survey characteristics, and the interstellar extinction influence the results of our search. We discuss here how it affects our detection efficiency analysis.



\subsubsection{Influence of the interstellar extinction}
\label{detec_effi_dust}
Without dust extinction along our lines of sight, RCB stars closer than $\sim16$, $\sim12$, $\sim8.8$ and $\sim5.8$ kpc for RCB absolute magnitudes $M_V$ of -5, -4, -3 and -2, respectively, would not be detected by EROS-2, due to saturation of the detectors, assuming a distance modulus of 14.4 for the Galactic centre, corresponding to $R_0 = 7.6\pm0.3$ kpc \citep{2005ApJ...628..246E}. In our fields at low Galactic latitude ($\vert b \vert < 8\degr$, see Fig. \ref{cgfields}), the concentration of dust is high, and the photometry of the new RCB stars is most likely affected by extinction. As extinction increases with decreasing Galactic latitude, their detection depends on the field monitored.

As most known RCB stars have $-5<M_V<-3$, most RCB candidates in the Bulge would thus be saturated (and therefore undetected) on EROS-2 images unless extinction is significant. However, our fields with $\vert b \vert<5\degr$ have sufficient absorption to ensure that Bulge RCB's with $M_V>-5$ are not saturated (see Sect.~\ref{subsec_Extinction}). Our detection efficiency is therefore independent of the absolute magnitude of Bulge RCBs in these fields, which constitute about 75\% ($\sim58$ deg$^2$, 61 fields) of the total Bulge area monitored.

For $\vert b \vert>5\degr$, bright RCBs would be saturated, but only 25\% of the EROS-2 fields are in this danger zone. We note the possibility that these bright RCBs could not be saturated on our reference images if they are in a declining phase, but it is less likely as the reference images are build with 15 images, distributed in time. These 25\% fields will not be used to measure the RCB scale height in Section~\ref{distrib}. As the minimum extinction observed in those fields is $A_{B_E}\sim0.7$ mag, the detection loss due to saturation will not affect Bulge RCB stars fainter than $M_V\lesssim-3.7$. 

\subsubsection{Influence of the survey characteristics}

Two main factors have to be considered in estimating the proportion of RCB stars that are detectable in our survey. First, since the EROS-2 monitoring period of 6.7 years is finite, RCB stars can remain bright or faint throughout the entire observing period and thus be missed by our detection algorithm, which was targeting only large luminosity drops ($>2.5$ mag). The evaluation of this effect is complicated by the fact that the frequency of fadings varies from one RCB to another by a factor $\sim100$ \citep[see][]{1996AcA....46..325J}, this frequency being strongly correlated to the $\left[C/H\right]$ abundance ratio. Second, potential luminosity drops can also remain undetected because our light curve analysis is affected by the observational sampling. We tried to evaluate the effect of these factors by studying closely the behaviour of known RCB stars and by simulating synthetic RCB light curves. The details of the studies are as follows.

\begin{itemize}
\item We need to estimate the average probability that an RCB star would not have a luminosity drop larger than 2.5 magnitudes during the 6.7-year observing period.  We know that 3 of the 5 known Galactic RCB stars located in our fields ($60\%$) remained bright during that period (see Sect.~\ref{knownRCB}). However, we found a larger value of $\sim94\%$ by studying the AAVSO light curves of two well known RCB stars, R CrB and RY Sgr, which have been monitored during more than a century with a median sampling of $\sim 0.3$ day. This last value corresponds to the probability for a survey with a time window identical to that of EROS-2 to not encounter a drop in brightness of more than 2.5 mag.  But as those two RCB stars might not be representative, but have different dust production rates and thus drop in luminosity with different frequencies, we decided to use the value estimated by \citet{2001ApJ...554..298A} based on a larger sample. They found than only 23 of 31 known Galactic RCB stars monitored by AAVSO had experienced at least one decline deeper than 3 mag during the 7 years of their observation. It corresponds to a probability of $\sim75\%$, which is also an intermediate value between the two previous ones.

\item RCB stars could also stay non-detectable in a faint phase during all the EROS-2 observation, but it is difficult to estimate the probability of this occurring. We know that R CrB and RY Sgr have never shown a faint phase that last longer than 3.3 years. However, most of the known RCBs could be only the tip of the iceberg, as higher dust production could be a more general case. For example, the SMC star MSX-SMC-014 was recently found to be an RCB candidate by \citep{2005ApJ...631L.147K} from its infrared spectral characteristics, but we found no optical counterpart in the EROS-2 database and also no brightness variation during the 6.7 years observation (EROS-2 SMC magnitude limit : $B_E>22.4$ and $R_E>21.7$). Another good example is the RCB star V854 Cen. It is the 3rd brightest RCB star in the sky at present but it was generally fainter than 13th magnitude and no brighter than 10th from 1913 to 1952 \citep{1996PASP..108..225C}.

\item If an RCB star is too faint on our reference images, it will not be catalogued and thus detected. As 15 images, distributed in time, were used to build the reference image, the chance of missing an RCB star at that step is low. Of the 17 LMC RCB stars listed by \citet{2001ApJ...554..298A}, 16 were located in our fields and all were catalogued. On the 6 known Galactic RCB stars, only one was too faint (see Sect.~\ref{knownRCB}). Therefore, from those 22 known RCB stars we estimate our efficiency at the catalogue step to be $\sim95\%$.


\item The number of measurements in the light curve of a star varies depending on its position on the CCD. Stars located on the edge have fewer measurements due to the pointing accuracy.

To evaluate the influence of the real observational sampling on our detection efficiency, we used a Monte Carlo simulation light curves. We added 1 to 3 brightness drops to simulated light curves of bright stars ($B_E=13$) with epochs identical to the fields and distributed in each of the monitored sky areas. Each luminosity drop imitates those observed in known RCB stars and has three consecutive phases: A steep decline at a constant rate of +0.045 mag.day$^{-1}$, corresponding to the average for all the new Galactic RCBs, see Tab.~\ref{tab.RCB.MagInfo}); then a plateau; and finally a recovery to the original magnitude at a rate of -0.02 mag.day$^{-1}$, slower than the  rate of the decline. The simulation has three randomly generated parameters: The date at the start of the decline (chosen to be during the EROS-2 observations); the duration of the plateau (from 10 to 300 days); and its depth (from 2.5 to 8 mag). The light curves created were then subjected to the same search pipeline as the real ones. The effect of the last selection criteria, the visual examination, was estimated by looking at one hundred simulated light curves, randomly selected. We estimate that the visual selection could not affect the final detection efficiency by more than 5\%. Overall, this simple simulation showed that our detection efficiency for obscured RCB stars is excellent. On average, it exceeds $94\%$ for the 112 main sky areas monitored (Galactic centre and spiral arms) and drops to $\sim45\%$ for the 18 less frequently observed fields (they are shown on Figure~\ref{cgfields}).
\end{itemize}

In summary, the detection of RCB stars should be of the order of $\sim60\%$ ($0.75\times0.95\times0.94$) complete for the  majority of our survey ($\sim 80\%$). Some infrequently observed fields may be less complete in RCB star detection.


\subsection{Visual inspection of pre-selected catalogued stars}
\label{vis_inspec_sect}
As a check of the light curve analysis, we inspected the light curves of all stars in our EROS-2 fields that match two characteristics of RCB stars: A carbon star spectrum and an infrared excess.

Therefore, we inspected the light curves of all stars listed in the carbon star catalogue \citep{2001BaltA..10....1A} and showing an infrared excess ($J-K>1$), the Bulge carbon stars found by \citet{1991A&AS...88..265A} and objects in the 2MASS catalogue with a high infrared excess (see the selection area delimited on Fig.~\ref{infra_red}). 35 carbon stars and 2082 2MASS objects (with an EROS-2 matched object within 2 arcsec) were inspected. The visual inspection resulted in the rediscovery of 10 RCBs already detected by the light curve analysis, but no new RCB stars appeared. This supports the good efficiency of our first analysis. The re-discovered RCBs show a high infrared excess in Figure~\ref{infra_red} since the 2MASS epochs by chance happened for nine of them to be either in a faint or declining/recovering phase.

\subsection{Spectroscopic selection}
\label{spect_sect}
If a well-sampled light curve is available, identification with the RCB class can be made with fairly high confidence because of the distinct nature of the RCB brightness drops. Of the 23 candidates, half showed a drop of less than 4 magnitudes and were thus not strong RCB candidates. Spectroscopic information was necessary to reveal their real nature.

\begin{figure*}
\centering
\resizebox{\hsize}{!}{\includegraphics{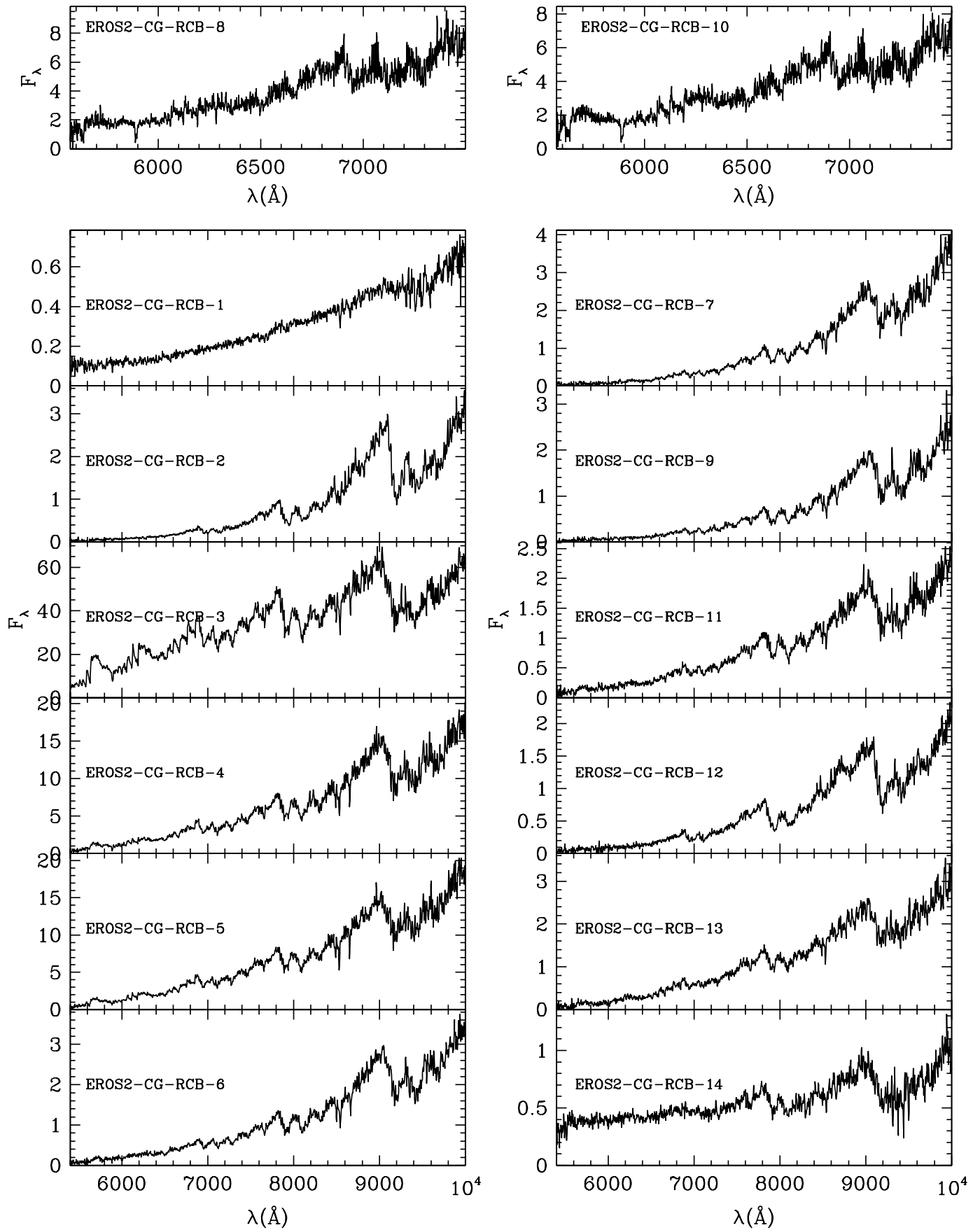}}
\caption{Spectra of the RCB star candidates. $F_{\lambda}$ is in units of 10$^{-15}$ erg cm$^{-2}$ s$^{-1}$ \AA$^{-1}$. The top two spectra have a resolution of 2 \AA{} while the bottom 12 spectra have a resolution of 7.5 \AA.  No correction for interstellar reddening has been applied.}
\label{spectrum_fig}
\end{figure*}

Twenty one candidates were observed during the nights JD-2454215 and JD-2454348; the last two were too faint (cg0334k19434, cg6271l19693). Of the 21, 13 have spectra with carbon features due to C$_2$ and/or CN molecules (see Fig.\ref{spectrum_fig}) and are therefore considered as confirmed new R Coronae Borealis stars. In addition to these 13 stars, we retain one other RCB candidate (cg0030l3740), because its light curve has multiple declines of more than 8 magnitudes. The spectrum of this star (Fig.~\ref{spectrum_fig}) is almost totally featureless, although some very weak CN bands are present ($\lambda\sim7876\AA{}$ and $9100\AA{}$). The star was very faint during the observation. The identification numbers, positions, and alternative names for our final sample of 14 new RCB star are given in Table~\ref{tab.RCB.GenInfo}.


One of these 14 new RCB stars, EROS2-RCB-CG-2, has a light curve that resembles that of the unusual Galactic RCB star DY Per \citep{1994BaltA...3..410A}. Its light curve shows declines of $>4$ mag. that are slower than those of most RCB stars. The declines are also associated with more symmetric recoveries.


The remaining 7 stars show M-type spectra, corresponding to oxygen-rich red giants. They were therefore rejected as RCB stars, even though their light curves show sudden drops in luminosity by up to 3.5 magnitude. In addition, we did not keep the 2 objects that were too faint for spectroscopy: The star cg6271l19693 has a light curve that resembles those of the M-type red giants, and cg0334k19434 has a light curve which is more like that of an AM Her star. The identifications and positions of these 9 stars are listed in Table~\ref{tab.RejectedCand}.

\begin{table*}
\caption{General information on the new Galactic RCB stars.
\label{tab.RCB.GenInfo}}
\medskip
\centering
\begin{tabular}{lllll}
\hline
\hline
EROS2 Galactic & EROS2 star & Coordinates ($J_{2000}$) & Other Id & General information \\
RCB name & Identifier & & & \\
\hline
\object{EROS2-CG-RCB-1}$^\star$ & \object{cg0030l3740}   &  17:52:19.96 -29:03:30.8  & &        \\
\object{EROS2-CG-RCB-2}$^\star$ & \object{cg0030m154}    &  17:52:48.70 -28:45:18.9  & & DY Per candidate     \\
\object{EROS2-CG-RCB-3}$^\star$ & \object{cg0062l18324}  &  17:58:28.27 -30:51:16.4  & & just outside the OGLE2 field SC23 \\
\object{EROS2-CG-RCB-4}$^\star$ & \object{cg0711m12518}  &  17:46:16.20 -32:57:40.9  & \object{Terz V 2046}$^{a}$ &        \\
\object{EROS2-CG-RCB-5}$^\star$ & \object{cg0715n14430}  &  17:46:00.32 -33:47:56.6  & &        \\
\object{EROS2-CG-RCB-6}$^\star$ & \object{cg1024m18795}  &  17:30:23.83 -30:08:28.3  & \object{V1135 Sco}$^{b}$ &        \\
\object{EROS2-CG-RCB-7}$^\star$ & \object{cg1026l13692}  &  17:29:37.09 -30:39:36.7  & \object{Terz V 1680}$^{a}$ &        \\
\object{EROS2-CG-RCB-8}$^{\star\star}$ & \object{cg1066m15039}  &  17:39:20.72 -27:57:22.4  & \object{Terz V 2960}$^{c}$ &        \\
\object{EROS2-CG-RCB-9}$^\star$ & \object{cg1127l13076}  &  17:48:30.87 -24:22:56.5  & &        \\
\object{EROS2-CG-RCB-10}$^{\star\star}$ & \object{cg1133k13881}  &  17:45:31.41 -23:32:24.4  & &        \\
\object{EROS2-CG-RCB-11}$^\star$ & \object{cg1187l16577}  &  17:48:41.53 -23:00:26.5  & & just outside the OGLE2 field SC15 \\
\object{EROS2-CG-RCB-12}$^\star$ & \object{cg1235m14181}  &  17:19:58.50 -30:04:21.3  & &        \\
\object{EROS2-CG-RCB-13}$^\star$ & \object{cg6110k17996}  &  18:01:58.22 -27:36:48.3  & \object{MACHO-176.19607.1138} &        \\
\object{EROS2-CG-RCB-14}$^\star$ & \object{cg6267k28863}  &  18:13:14.86 -27:49:40.9  & &        \\
\hline
\multicolumn{5}{l}{$^{a}$ \citet{1988A&AS...76..205T}, $^{b}$ \citet{1971GCVS3.C......0K}, $^{c}$
 \citet{1991A&AS...90..451T}}\\
\multicolumn{5}{l}{$^\star$ Spectral resolution 7.5 \AA{} per pixel. $^{\star\star}$ Resolution 2.0 \AA{} per pixel.} \\
\end{tabular}
\end{table*}

\begin{table*} 
\caption{General information on the rejected candidates.
\label{tab.RejectedCand}}
\medskip
\centering
\begin{tabular}{lllll}
\hline
\hline
EROS2 star & Coordinates ($J_{2000}$) & Classification & Other Id & General information \\
Identifier & & & & \\
\hline
\object{cg0334k19434}   & 18:18:15.10 -31:51:54.6 & AM Her & \object{MACHO-155.26433.492} & \\
\object{cg0606k27985}$^{\star\star}$  & 18:16:54.22 -25:02:54.7 & M5 & \object{MACHO-177.26016.56}  &  \\
\object{cg1047k14026}$^\star$  & 17:28:07.82 -29:48:14.0 & M5 &   & \\
\object{cg1076k13424}$^\star$  & 17:33:26.21 -27:32:02.7 & M2 &  & suddenly appears 3 mag brighter \\
\object{cg1077m26188}$^\star$  & 17:36:15.13 -27:37:11.4 & M6 &  & periodic? \\
\object{cg1131n13463}$^{\star\star}$  & 17:46:00.16 -23:21:16.1  & M0 & \object{ASAS174600-2321.3}$^{a}$  & drop of 2.5 mag in ASAS light curve \\
\object{cg1135m3139}$^\star$   & 17:46:12.87 -23:49:44.2 & M3 &   & \\
\object{cg6126m5824}$^\star$   & 18:05:30.42 -31:53:24.2 & M2 &  &  \\
\object{cg6271l19693}  & 18:13:16.97  -25:31:34.8  & M? &   &        \\
\hline
\multicolumn{5}{l}{$^{a}$ ASAS: All Sky Automated Survey \citep{1997AcA....47..467P}} \\
\multicolumn{5}{l}{$^\star$ Spectral resolution 7.5 \AA{} per pixel. $^{\star\star}$ Resolution 2.0 \AA{} per pixel.} \\
\end{tabular}
\end{table*}

\subsubsection{Radial velocities of confirmed RCBs}

Radial velocities were obtained by cross-correlating spectra obtained at resolutions of 17, 45, 72 or 151 km/s/pixel during the night JD-2454348, using the spectrum of the carbon star X Vel as a template. The radial velocity of X Vel was taken to be -5.4 km/s (determined from high resolution echelle spectra which were cross-correlated against the radial velocity standard $\alpha$ Cet with a velocity of -25.8 km/s). The heliocentric radial velocities and their errors are listed in Table~\ref{tab.RadialVel}.  We did not measure the radial velocity of RCB-1 as its spectrum shows no carbon features. 

The measured radial velocities cover a large range, from -325 to 115 km s$^{-1}$, with no convincing trend with Galactic latitude. This large range of values is consistent with the radial velocities found for OH/IR stars, oxygen-rich evolved stars, located in the Galactic Bulge at that particular Galactic longitude \citep[see][fig.12]{1999MNRAS.310..629S}.


\begin{table} 
\caption{Heliocentric radial velocities and $^{13}$C presence
\label{tab.RadialVel}}
\medskip
\centering
\begin{tabular}{lccccc}
\hline
\hline
RCB  & l & b & $V_{hel}$ & $\sigma_V$ & $^{13}$C \\
     & (deg) & (deg) & $(km s^{-1})$ & $(km s^{-1})$ & presence?\\
\hline
1 & 0.65107 & -1.32119 & - & - & -\\
2 & 0.96560 & -1.25721 & 68.4 & 13.7 & -\\
3 & -0.23313 & -3.37369 & 46.9 & 2.6  & no\\
4 & -3.3663 & -2.21249 & -246.4 & 3.1 & no\\
5 & -4.11175 & -2.59990 & -323.1 & 0.7 & no\\
6 & -2.79018 & 2.15220 & 114.2 & 3.5  & no\\
7 & -3.31629 & 2.00684 & -111.4 & 6.8  & -\\
8 & 0.10674 & 1.68854 & 111.9 & 3.7  & no\\
9 & 4.23037 & 1.80687 & 112.5 & 1.0  & -\\
10 & 4.59793 & 2.82737 & 76.4 & 3.8  & no\\
11 & 5.43103 & 2.47977 & 73.0 & 4.4  & no\\
12 & -3.99599 & 4.05891 & -43.7 & 5.4  & -\\
13 & 2.96514 & -2.43305 & -292.1 & 0.9  & -\\
14 & 3.98548 & -4.72163 & -11.4 & 0.8  & -\\
\hline
\end{tabular}
\end{table}

\subsubsection{$^{13}$C test and Ca II triplet}
\label{13C_CaII}
The absence of enhanced $^{13}$C is a characteristic of RCB stars \citep{1994MNRAS.268..544P}. We tried to estimate the abundance of $^{13}$C using $^{13}$CN band at 6260 $\AA{}$ \citep[see][]{1991MNRAS.249..409L}. It was possible to determine the absence of a significative presence of $^{13}$C for seven RCBs of the 14 new RCBs. The spectra of these 7 RCBs are higher resolution ($2\AA{}$). For the remaining RCBs, the low resolution spectra and the small signal due to extinction, prevent us from determining the abundance of $^{13}$C. The results are summarized in Table~\ref{tab.RadialVel}.
 
An interesting feature in the spectrum of warm carbon stars is the infrared triplet due to ionised Calcium ($\lambda\sim8498$, 8543 and 8662 $\AA{}$). \citet{1971ApJ...167..521R} shows that the intensity of those lines is a good indicator of the carbon star temperature; the cooler the temperature, the weaker the lines. In the spectra presented in Figure~\ref{spectrum_fig}, the Ca II feature could be observable only in the case of low resolution spectra. On these 12 spectra, 10 present either strong or moderate Ca II lines and two have no or very weak Ca II lines (EROS2-CG-RCB-2 and -12). These 2 RCBs are therefore intrinsically cooler than the others. We note that EROS2-CG-RCB-2 is a DY Per RCB, which are known to be cool RCBs. This remark is important as all spectra are strongly reddened, the stars being close to the Galactic plane. This temperature indication will be use in Section~\ref{subsec_Extinction} as a check of our extinction correction.

\section{The new R Coronae Borealis stars}
\label{sec_NewRCB} 

None of the newly discovered RCB stars in Table~\ref{tab.RCB.GenInfo} is located in the spiral arm fields; they are all found in the direction of the Galactic Bulge, distributed equally between the north and south sides of the Galactic plane. Their positions are shown on the Bulge field map in Figure~\ref{cgfields}, finding charts in Figure~\ref{charts}, and the light curves in Figures~\ref{lc} -- \ref{lc_end}. The actual measurements are available at \texttt{http://eros.in2p3.fr/Variables/RCB/RCB-CGBS.html} .

Only one of the 14 new RCB stars, EROS2-CG-RCB-13, is located in the MACHO fields (MACHO identifier 176.19607.1138). None is located in the OGLE-II fields.
\begin{table*} 
\caption{Near-IR photometry.
\label{tab.RCB.IRInfo}}
\medskip
\centering
\begin{tabular}{llrrrlrrr}
\hline
\hline
RCB name & JD Epoch 2MASS & $J_{\mathrm{2MASS}}$ & $H_{\mathrm{2MASS}}$ & $K_{\mathrm{2MASS}}$ & JD Epoch Denis & $I_{\mathrm{DENIS}}$ & $J_{\mathrm{DENIS}}$ & $K_{\mathrm{DENIS}}$  \\
\hline
EROS2-CG-RCB-1 & 2451010.6571$^\star$ & 12.502 & 10.845 & 9.077 &  2451404.57740$^\star$ & 14.582 & 13.635 & 9.615 \\
EROS2-CG-RCB-2 & 2451010.6591$^\vee$ & 11.389 & 10.061 & 9.290 &  2451410.59416$^\wedge$ & 14.394 & 11.223 & 9.197 \\
EROS2-CG-RCB-3 & 2451040.5447$^\star$ & 12.702 & 10.348 & 8.538  &   &  &  &  \\
EROS2-CG-RCB-4 & 2451039.6137$^\Diamond$ & 10.312 & 9.087 & 7.936  &   &  &  &  \\
EROS2-CG-RCB-5 & 2451039.6095$^\star$ & 15.524 & 12.715 & 9.977   &   &  &  &  \\
EROS2-CG-RCB-6 & 2451035.6283$^\Diamond$ & 10.193 & 9.151 & 8.277  &  2450313.53086$^?$ &  & 13.882 & 9.487 \\
EROS2-CG-RCB-7 & 2451035.6272$^\vee$ & 11.806 & 10.310 & 9.061  &  2450216.79936$^?$ &  & 14.755 & 10.494 \\
		&		&	&	&	 &  2450313.53012$^?$ &  & 14.264 & 9.856 \\
EROS2-CG-RCB-8 & 2450996.7036$^\star$ & 13.152 & 11.518 & 9.895  &   &  &  &  \\
EROS2-CG-RCB-9 & 2451010.6160$^\vee$ & 12.458 & 10.484 & 8.808  &  2451074.74752$^\vee$ & 16.633 & 12.509 & 8.854 \\
EROS2-CG-RCB-10 & 2450963.8322$^\wedge$ & 11.117 & 9.534 & 7.975 &  2451103.66233$^\Diamond$ &  & 14.878 & 9.202 \\
EROS2-CG-RCB-11 & 2450950.8589$^\Diamond$ & 10.106 & 9.059 & 8.109  &  2451074.74555$^\Diamond$ & 12.535 & 10.345 & 8.183 \\
EROS2-CG-RCB-12 & 2451035.5272$^\star$ & 13.848 & 11.949 & 10.425  &  2451381.63688$^\star$ & 16.687 & 12.951 & 9.876 \\
EROS2-CG-RCB-13 & 2451364.6757$^\star$ & 13.203 & 11.330 & 9.490  &  2450174.90133$^\star$ & 15.374 & 13.393 & 9.699 \\
EROS2-CG-RCB-14 & 2451749.4873$^\Diamond$ & 11.347 & 9.807 & 8.340  &   &  &  &  \\
\hline
\multicolumn{9}{l}{$\star$ : during a faint phase, $\Diamond$ : during a bright phase, $\vee$ and $\wedge$ : during a dimming or recovering phase, ? : phase unknown. }\\
\end{tabular}
\end{table*}
\subsection{Infrared properties}

\begin{figure} 
\centering
\includegraphics[width=8.25cm]{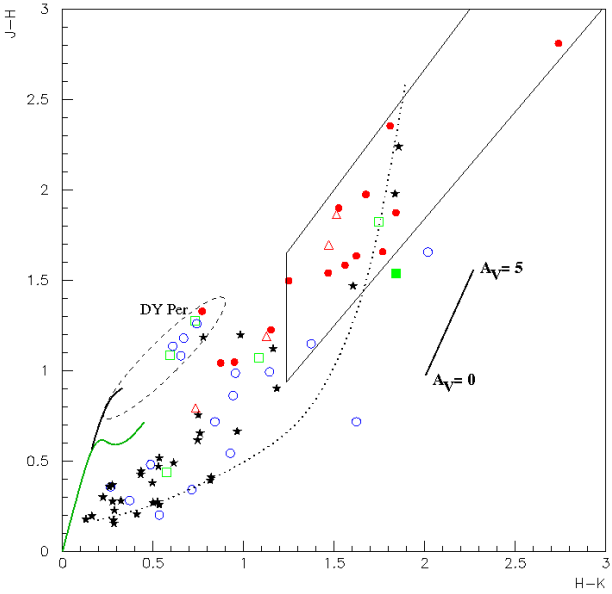}
\caption{The $J-H$ versus $H-K$ colour diagram. All the known RCB stars, including those of DY Per type, are represented. The black stars are the 31 confirmed Galactic RCBs listed by \citep{1996PASP..108..225C} plus ES Aql and V2552 Oph. The 4 open red triangles are the ones recently found by \citet{2005AJ....130.2293Z}. The blue circles are the 17 RCB stars found in the LMC \citep{2001ApJ...554..298A}. The green open squares are the 5 RCBs found in the SMC \citep{2004A&A...424..245T} and the full green square is the SMC RCB star MSX-SMC-014 \citep{2005ApJ...631L.147K}. The 14 new EROS-2 Galactic RCB stars are indicated with red dots. The location of the DY Per stars is delimited by the dashed ellipse, which also encloses the area occupied by most carbon-rich stars. The dotted curve corresponds to the combination of blackbodies consisting of a 5500 K star and a 1000 K dust shell in various proportions ranging from all 'star' to all 'shell' \citep[from][]{1997MNRAS.285..339F}. The area delimited by straight lines shows the zone in the 2MASS database from which stars in the EROS-2 fields were selected for light curve examination (see Sect.~\ref{vis_inspec_sect}). The line on the right side represents the reddening vector from \citet{1985ApJ...288..618R}. Also shown are the expected position (lines in the bottom-left side) for common dwarf (green) and giant (black) stars from \citet{1988PASP..100.1134B}.}
\label{infra_red}
\end{figure}

Most RCB stars have a near-IR excess even if they have not shown a decline for several years. Dust can still form continuously near the star, without completely obscuring its photosphere \citep{1997MNRAS.285..339F}. JHK measurements have been made by the 2MASS project for all the new Galactic RCBs (see Tab.~\ref{tab.RCB.IRInfo}), and all have $J-K>1.9$, indicative of an infrared excess (see Fig.~\ref{infra_red}). Part of this excess is due to the high interstellar extinction, but we note also that the 2MASS epochs for 10 of the 14 RCBs coincide with a faint or dimming/recovering phase.

EROS2-CG-RCB-2 is in the region of Figure~\ref{infra_red} occupied by common carbon stars: this area also corresponds to the typical position for DY Per-type RCB stars. This result, along with the DY Per-like light curve, confirms the membership of this star in the DY Per class of RCB stars.


We note that all new RCB stars have entries in the MSX6C Infrared Point Source Catalog \citep{2003yCat.5114....0E}, except the DY Per-like RCB EROS2-CG-RCB-2 and EROS2-CG-RCB-12.


\subsection{Extinction correction and distances}
\label{subsec_Extinction}

\begin{figure} 
\resizebox{\hsize}{!}{\includegraphics{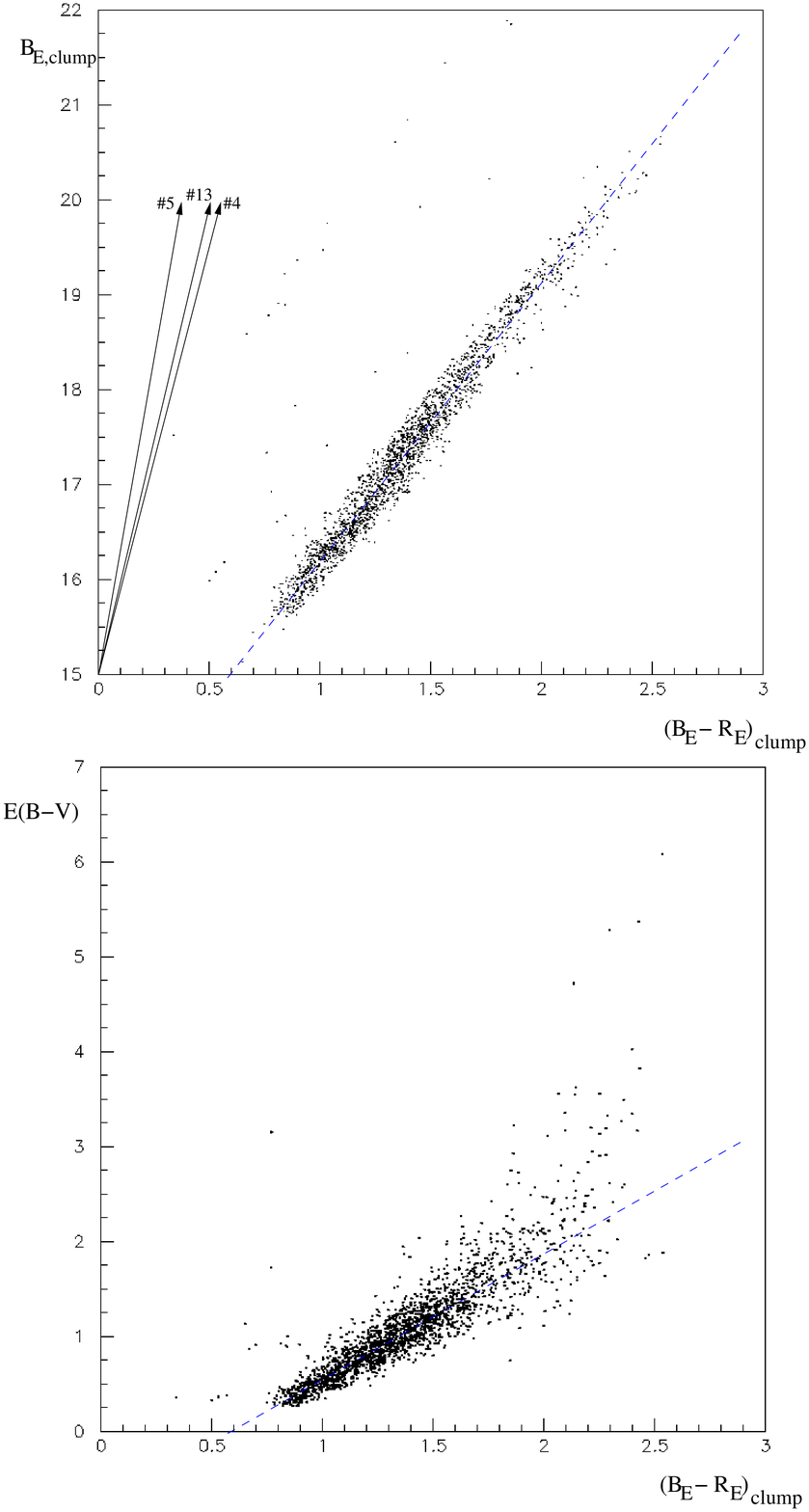}}
\caption{Top panel: Apparent magnitude $B_{E,clump}$ vs. $(B_E - R_E)_{clump}$ for the centre of the red giant clump in each quarter-CCD subfield. The expected linear relation due to reddening can be observed. The vectors on the left side represent the absorption due to carbon cloud, estimated for three RCBs during a decline. Bottom panel: the reddening $E(B-V)$ to the Bulge derived from the COBE/DIRBE data by \citet{1998ApJ...500..525S} versus $(B_E - R_E)_{clump}$.}
\label{schlegelclumpcolor}
\end{figure}

If all new RCBs are in the Bulge, then their apparent luminosity is strongly affected by reddening. The extinction towards small areas of the Bulge can be determined from the position of the red-giant clump in the colour-magnitude diagram. The selection of clump-giant samples was performed for each quarter-CCD subfield (about 10.3'x10.3') as described in \citet{2006A&A...454..185H}. The result is presented in the top panel of Figure~\ref{schlegelclumpcolor}, where the expected linear dependence between the fitted magnitudes $B_{E,clump}$ and the colours $\left\langle B_E - R_E \right\rangle_{clump}$ of the clump centre for different samples is shown. The all-sky COBE/DIRBE extinction map by \citet{1998ApJ...500..525S} was not used to estimate reddening because it overestimates the amount of dust in high-extinction areas \citep{1998astro.ph..2307S, 1999ApJ...512L.135A, 2003MNRAS.338..253D}. The lower panel of Figure~\ref{schlegelclumpcolor} confirms the non-linear dependence between the reddening $E(B-V)$ from \citet{1998ApJ...500..525S} and the reddening $\left\langle B_E - R_E \right\rangle_{clump}$ derived here. However, the lower panel of Figure~\ref{schlegelclumpcolor} can be used to define the EROS-2 zero-extinction colour of the Bulge red clump giant centre. We find $(B_E - R_E)_{0,clump} \approx 0.6$. Our intrinsic clump colour, when transformed by Equation~\ref{eq.photcalib}, is close to the expected colour $\left\langle V-I\right\rangle_{0,clump} \sim1.0$ for the giant clump in the Bulge \citep{2004MNRAS.349..193S}.

From Figure~\ref{schlegelclumpcolor}, we obtained the following visual extinction law:

\begin{equation}
$$ A_{B_E} = -1.77 + 2.95 \left\langle B_E - R_E\right\rangle_{clump}  $$.
\label{eq.A_BE}
\end{equation}

From this relation, we estimated the extinction to each RCB star based on the colour of clump-giant stars located within 3 arcmin. The correction applied to each RCB star colour and magnitude at maximum is given on Table~\ref{tab.RCB.MagInfo}. 

\begin{figure}
\resizebox{\hsize}{!}{\includegraphics{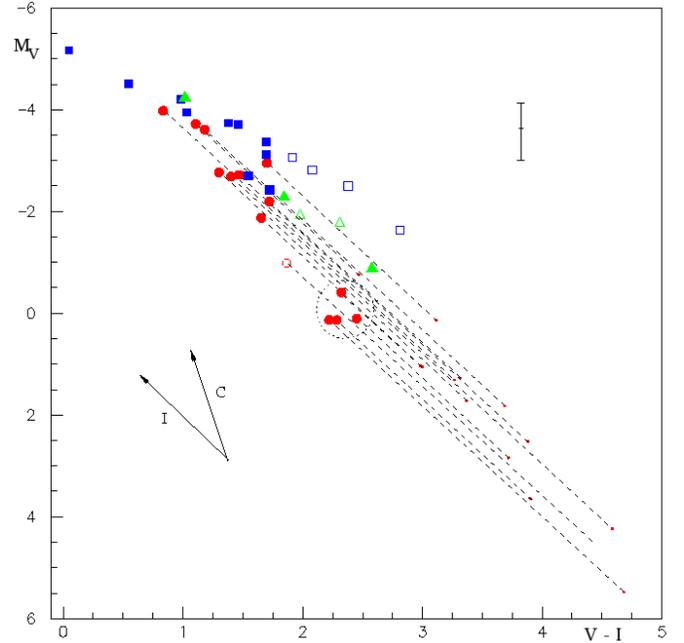}}
\caption{Absolute magnitude at maximum, $M_V$, vs. $V-I$; full symbols represent common RCB stars, open symbols the DY Per-type RCB stars. The red circles are the newly discovered Galactic RCBs, assuming a galactocentric distance, after the interstellar reddening corrections shown by a dashed line for each star. The location of 10 LMC RCBs, plus 4 DY Per-like RCBs, from \citet{2001ApJ...554..298A} are indicated with blue squares. 3 SMC RCBs, plus 2 DY Per-like RCBs, from \citet{2004A&A...424..245T} are shown by green triangles. All data are from the EROS-2 photometry. LMC and SMC RCB magnitudes have been corrected for their respective extinctions. The four Galactic RCBs surrounded by a dotted circle have an unusual position that can be explained by a thick circumstellar shell or a localisation behind the Bulge (see text). The vectors represent the reddening correction direction due to the interstellar medium (I) and a carbon shell (C) (see Fig.~\ref{schlegelclumpcolor}). The line in the up-right corner represents the lower and upper boundaries for a $\pm2$kpc Galactic Bulge radius.}
\label{vvsvi}
\end{figure}

\begin{figure*}
\includegraphics[scale=0.45]{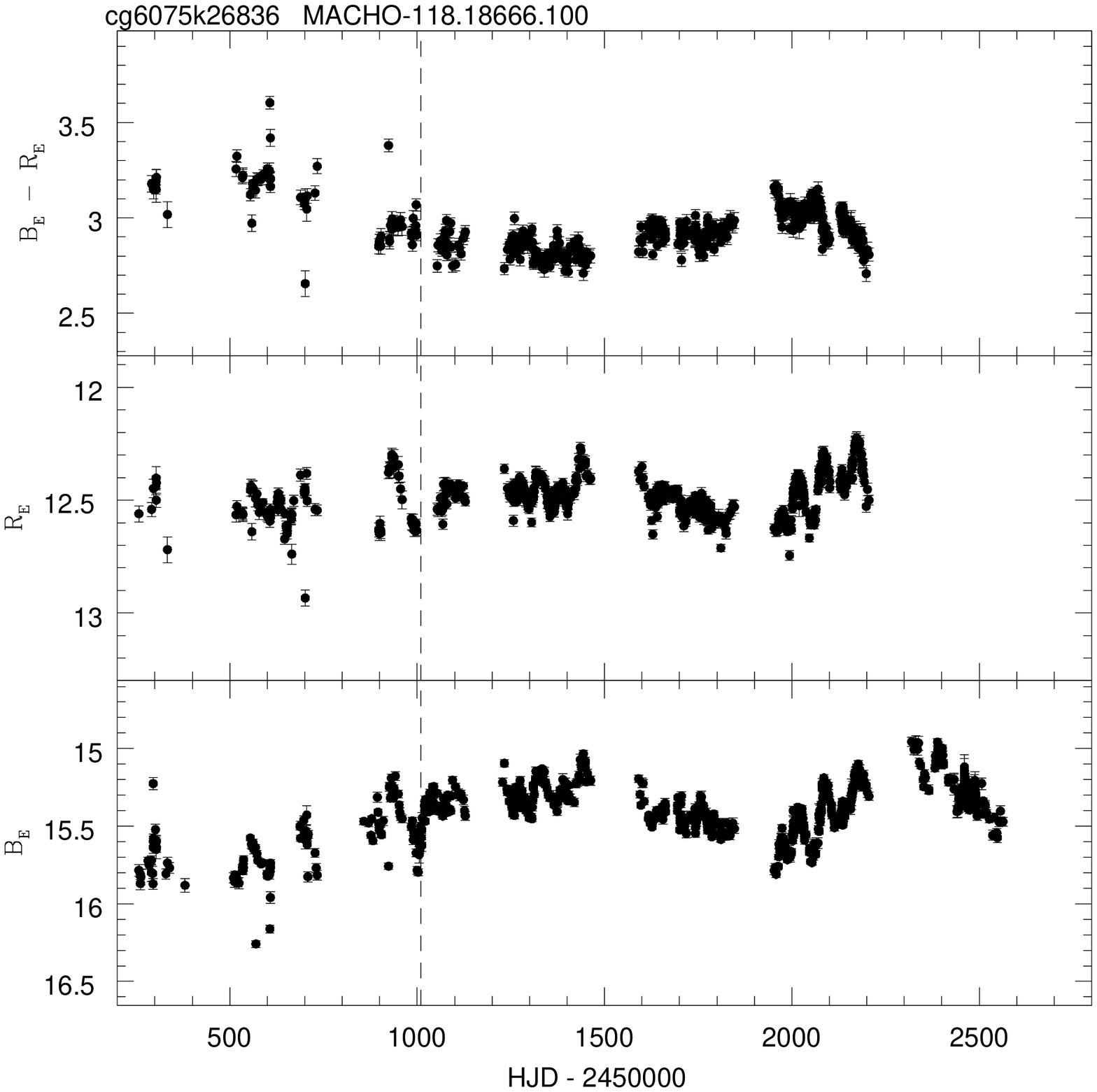}
\includegraphics[scale=0.45]{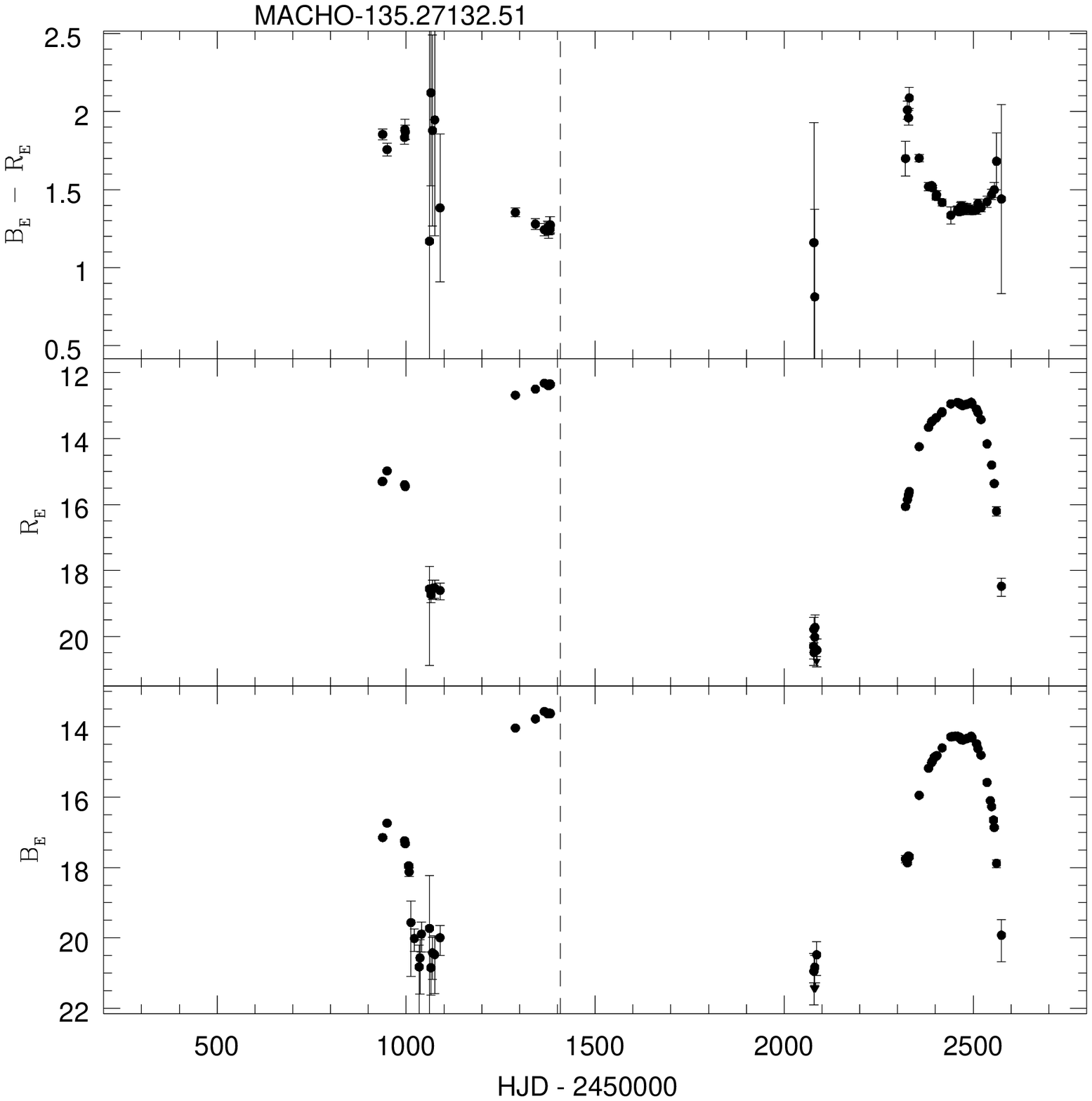}
\includegraphics[scale=0.45]{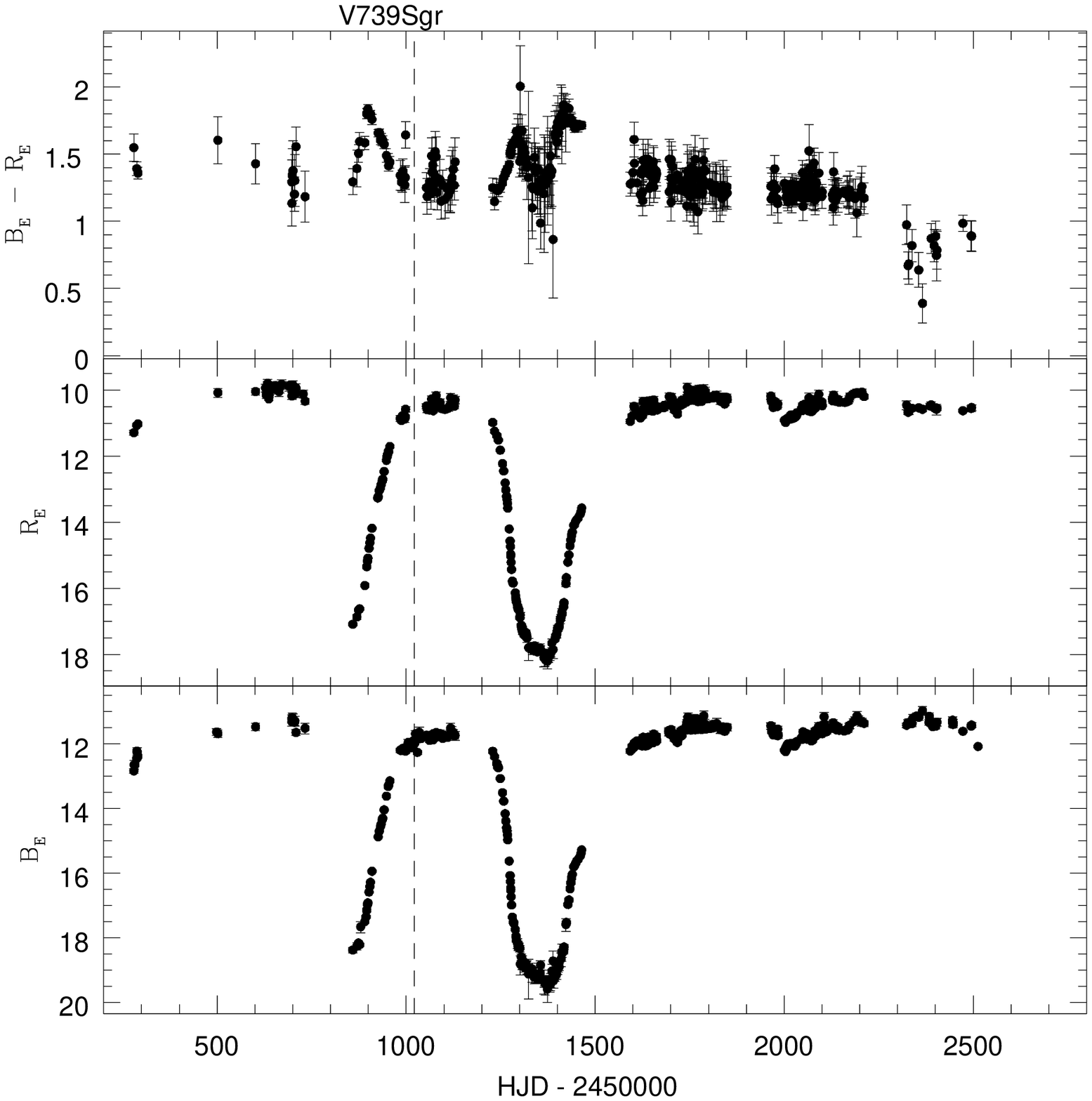}
\includegraphics[scale=0.45]{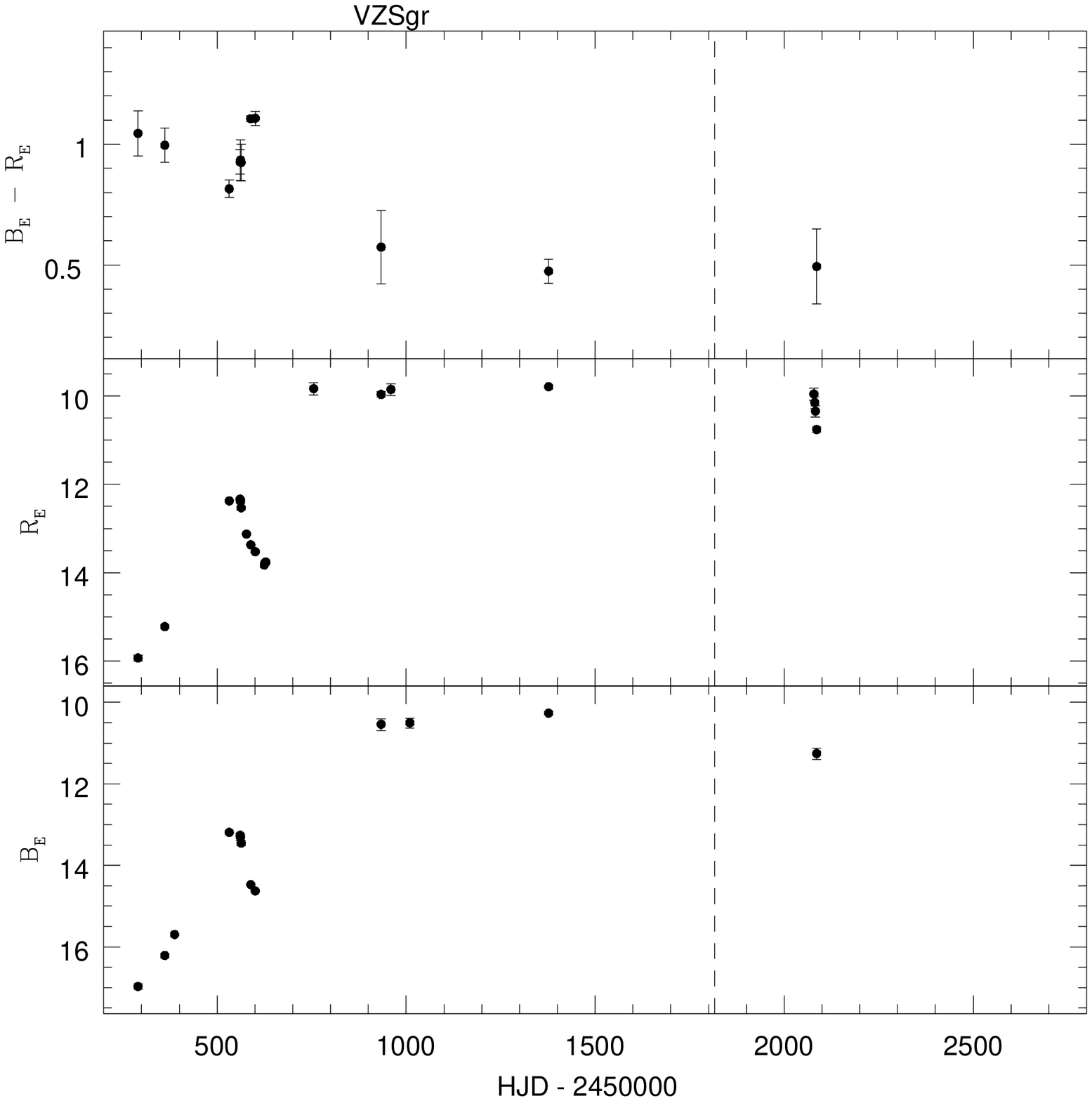}
\caption{EROS-2 light curves of MACHO-118.18666.100 (re-classified as a M giant star, see Sect.~\ref{knownRCB}) and three known Galactic RCB stars, MACHO-135.27132.51, V739 Sgr and VZ Sgr. The dashed vertical lines indicate the 2MASS epochs. The photometry of the last three RCB stars has been reprocessed as these stars were either saturated or too faint on the original EROS-2 reference images.}
\label{lc_KnownRCB}
\end{figure*}

The positions of the de-reddened RCB stars are shown in the $M_V$ vs. $V-I$ diagram, Figure~\ref{vvsvi}, and the derived intrinsic values, $M_V$ and ($V-I$)$_0$, are given in Table~\ref{tab.RCB.MagInfo}. The V and I magnitudes were obtained from $B_E$ and $R_E$ using Equation~\ref{eq.photcalib}. The positions of 10 (resp. 3) common RCBs\footnote{Three LMC RCB stars were not used: MACHO-6.6575.13 which never reached its maximum brightness in the EROS-2 data,  HV 12842 which is outside the EROS-2 fields, and the hot RCB star MACHO-11.8632.2507} and 4 (resp. 2) DY Per-like RCBs in the LMC (resp. SMC) \citep{2001ApJ...554..298A,2004A&A...424..245T} are also shown in Figure~\ref{vvsvi}. We used the EROS-2 light curve to measure the maximum brightness of each RCB star. The distance modulus used was 14.4 mag for the Galactic Centre (the present best measure of the galactocentric distance is $R_0 = 7.6\pm0.3$ kpc, see \citet{2005ApJ...628..246E}), 18.5 mag for the LMC and 18.9 mag for the SMC. We also corrected the LMC and SMC magnitudes for the total reddening (Galactic foreground + intrinsic dust), corresponding to $E(B-V)_{LMC}\sim0.17$ and $A_{V,LMC}\sim0.5$ mag for the LMC \citep{2001ApJ...554..298A}, $E(B-V)_{SMC}\sim0.06$ and $A_{V,SMC}\sim0.17$ mag for the SMC \citep{2002AJ....123..855Z}.


The fact that the Bulge reddening correction results in an overlap in Figure~\ref{vvsvi} between Galactic and Magellanic Cloud RCB stars suggests that most of the new RCB stars are indeed in the Galactic Bulge (the Bulge occupies a region between 6 and 10 kpc).  Note that without the reddening correction, the new Galactic RCB stars would be the coolest RCBs known ($T_{\rm eff}<3600 K$, with a median at $\sim$3000 K).  With the reddening correction, the effective temperatures of most of the new RCB stars are as expected for RCB stars. Most currently known RCB stars are carbon-rich supergiants with an effective temperature between 5000 and 7000 K \citep{1996PASP..108..225C}, which is true also for the RCB stars in the LMC.

\begin{table*}
\caption{Properties of the new Galactic RCB stars, including reddening estimates and derived absolute magnitudes and intrinsic colours.
\label{tab.RCB.MagInfo}}
\medskip
\centering
\begin{tabular}{lrcrrcrcrcrc}
\hline
\hline
RCB & $R_{E,max}$ & $(dR_E/dt)_{max}$ & Drop $R_E$ & $B_{E,max}$ & $(dB_E/dt)_{max}$ & Drop $B_E$ & $E(B-V)^{a}$ & $A_{B_E}$ & $\triangle(B_{E}-R_{E})$ & $M_V$ & ($V-I$)$_0$\\
Number & & $mag.day^{-1}$ & & & $mag.day^{-1}$ & & & & & &\\
\hline
1 & 12.44    & 0.038  	 & $>8.3$  & 14.40 	& 0.042   	& $>7.1$   & 1.42 &   4.31 & 1.46 & -3.98 &  0.83\\
2 & 14.48    & 0.021  	 & $>3.0$  & 17.10 	& 0.019   	& $>2.4$   & 2.56 &   4.43 & 1.50 & -0.98 &  1.87\\
3 & 11.17    & 0.037  	 & $>6.9$  & 12.65 	& 0.044   	& 7.4      & 1.26 &   2.42 & 0.82 & -3.73 &  1.10\\
4 & 12.54    & 0.067  	 & $>8.2$  & 14.75 	& 0.052   	& $>6.6$   & 2.03 &   4.43 & 1.50 & -3.61 &  1.18\\
5 & 13.52    & 0.102  	 & 6.0     & 15.75 	& 0.094   	& $>5.7$   & 1.38 &   2.24 & 0.76 &  0.09 &  2.45\\
6 & 12.75    & 0.034  	 & 6.2     & 14.77 	& 0.036   	& 6.0      & 2.40 &   3.66 & 1.24 & -2.77 &  1.30\\
7 & 14.05    & 0.033  	 & $>6.6$  & 16.80 	& 0.027   	& $>4.6$   & 3.82 &   5.63 & 1.91 & -2.67 &  1.40\\
8 & 13.04    & nd    	 & $>7.6$  & 15.37 	& nd     	& $>6.0$   & 1.39 &   4.28 & 1.45 & -2.72 &  1.47\\
9 & 15.20    & 0.033  	 & 2.6     & 18.01 	& 0.035   	& 1.8      & 3.14 &   4.37 & 1.48 &  0.13 &  2.22\\
10 & 12.46    & 0.060  	 & 6.0     & 14.26 	& 0.058   	& 5.4      & 0.98 &   2.40 & 0.81 & -1.88 &  1.65\\
11 & 12.35    & 0.058  	 & 5.8     & 14.34 	& 0.060   	& 5.1      & 1.14 &   2.83 & 0.96 & -2.20 &  1.72\\
12 & 14.14    & 0.025  	 & 3.2     & 16.48 	& 0.031   	& 3.9      & 1.06 &   2.86 & 0.97 &  0.13 &  2.28\\
13 & 11.42    & 0.043  	 & 5.1     & 13.29 	& 0.066   	& 6.2      & 1.38 &   2.51 & 0.85 & -2.94 &  1.70\\
14 & 12.45    & 0.026  	 & $>8.3$  & 14.24 	& 0.033   	& $>7.2$   & 0.49 &   1.18 & 0.40 & -0.41 &  2.32 \\
\hline
\multicolumn{10}{l}{nd = not detected,  $^a$ from \citet{1998ApJ...500..525S}}\\
\end{tabular}
\end{table*}

\begin{figure*}
\includegraphics[scale=0.45]{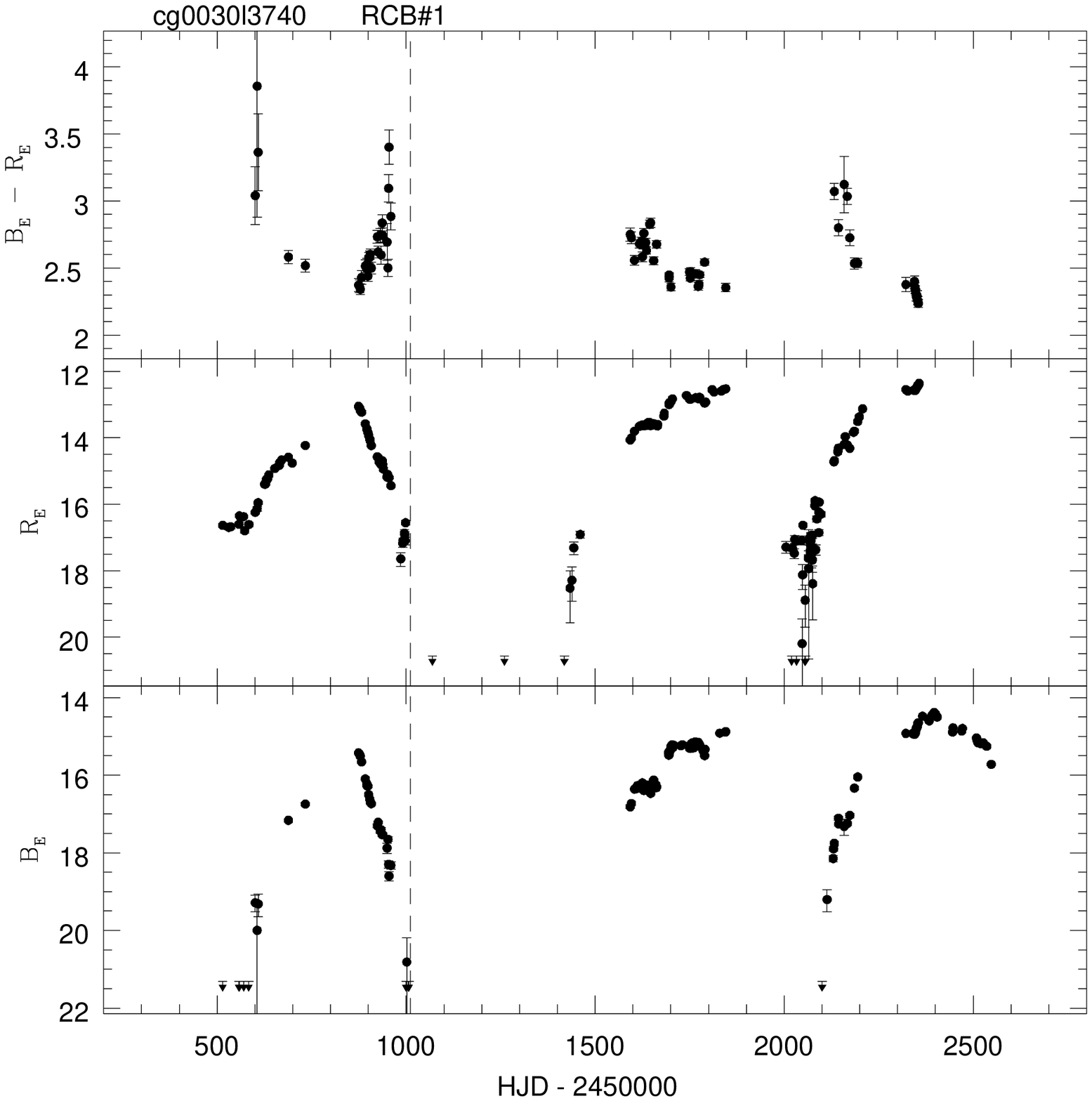}
\includegraphics[scale=0.45]{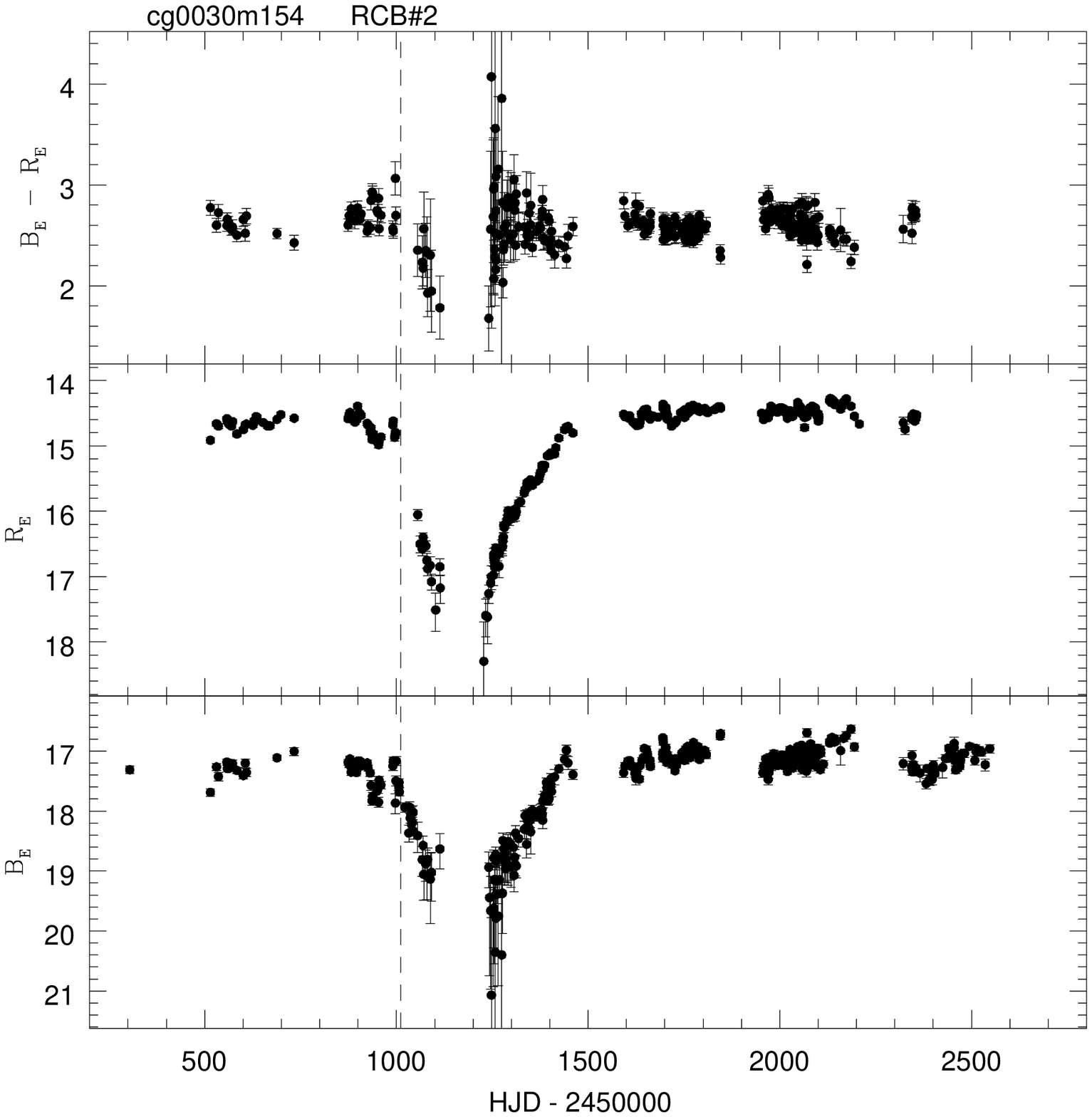}
\includegraphics[scale=0.45]{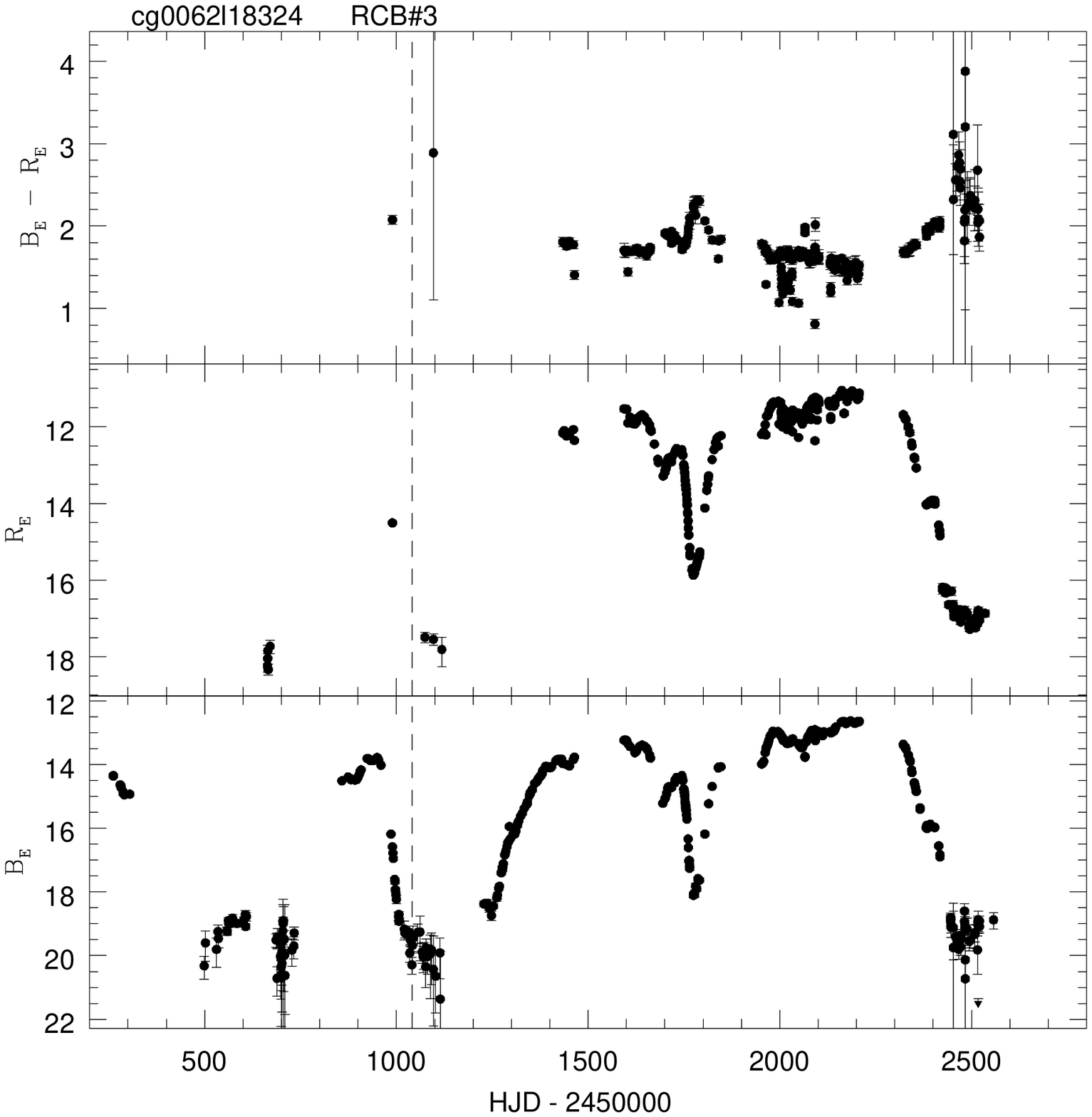}
\includegraphics[scale=0.45]{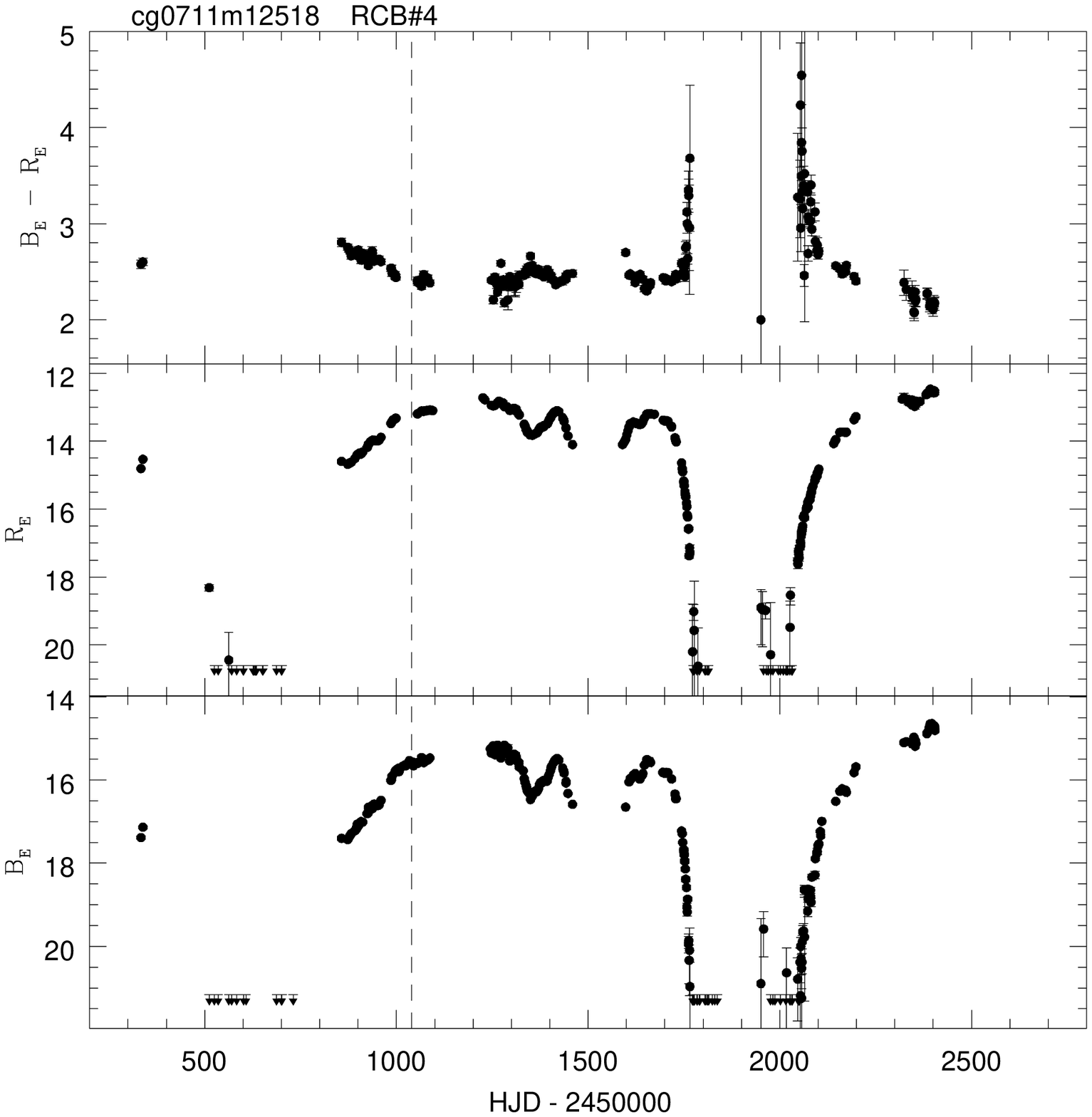}
\caption{Light curves of the 14 new RCBs star, one being a DY Per star (EROS2-RCB-CG-2). The arrows represent our detection limit. The dashed vertical lines indicate the 2MASS epochs. Top: $B_E - R_E$ colour vs. time; Middle: $R_E$ light curve ; Bottom: $B_E$ light curve.}
\label{lc}
\end{figure*}

Even with the Bulge extinction correction, four of the new Galactic RCBs (EROS2-CG-RCB-5, -9, -12 and -14) are cooler than known RCB stars ($T_{\rm eff}\sim3800$ K) and have unusually faint absolute magnitudes of $M_V\sim0$. For three of them (EROS2-CG-RCB-5, -9 and -14), we have an indication from the presence of strong Ca II triplet in their spectrum (see Sect.~\ref{13C_CaII}) that they are actually hotter with a temperature more likely similar to classical RCBs ($T_{eff}\sim5000$ K, which corresponds to an intrinsic V-I colour of $\sim1.2$). Therefore these 3 RCB stars are certainly subject to a supplementary extinction. We retain 3 hypothesis. First, they could be located beyond the Galactic Bulge at a maximum distance of $\sim14$ kpc\footnote{After reddening correction due to the supplementary extinction, $A_{V_{supp}}\sim2$ mag, the three RCBs are still located in the colour magnitude diagram (Fig.~\ref{vvsvi}) at about 1.3 mag below the LMC ones.}. We have thus assumed that the supplementary extinction comes from dust located beyond the Bulge. Second, these 3 RCBs may still be located inside the Bulge, but have circumstellar extinctions that are higher that those for known RCBs ($A_{V_{circum}}\sim3$ mag is needed). We would have thus not observed the maximum luminosity of these RCBs. The reddening vector due to a Carbon shell is different than the interstellar one: A factor 3.5 in the slope can be observed on Figure~\ref{schlegelclumpcolor}. The reddening vectors due to carbon clouds have been estimated for three of the new RCB stars during a decline\footnote{We selected only three RCBs based on the colour variation during a decline (see Fig.~\ref{lc} -- \ref{lc_end}): We did not retain some RCBs that become bluer due to a blend by a bluer star.}. Third, there may be a higher foreground extinction due to a compact interstellar dust clouds that happen to be in the line of sight\footnote{This can be observed in one case: The chart of the star EROS2-CG-RCB-5 (Fig.~\ref{charts}) shows a local high extinction area close to its position.}. However, we consider this possibility the least likely as the 3 Galactic RCBs would still be located in the colour magnitude diagram at about 1.3 mag below the LMC ones (if $T_{eff}\sim5000$ K), implying that their location is behind the Bulge without extra-reddening on the line of sight. We would thus be confronted with the improbable case where only the 3 RCB stars located beyond the Bulge in our new sample, would be affected by a compact foreground cloud on their line of sight. We note also that if the last hypothesis is true, we should observe the converse, some RCB stars that have been over-corrected due to a lower extinction than average. But it is not the case.

In the case of EROS-CG-RCB-12, we have already mentioned in Section~\ref{13C_CaII} that this star is intrinsically cooler than the other RCBs. Its position in Fig.~\ref{vvsvi} is therefore less surprising. It could be similar to the cool RCB star discovered in the SMC, EROS2-sm0067m28134b\footnote{We note that the SMC star EROS2-sm0067m28134b may not have reached its maximum brightness in the EROS-2 light curve, as the maximum observed lasted only a few days. If so, this cool SMC RCB star should also be brighter and bluer.} \citep{2004A&A...424..245T, 2005ApJ...631L.147K}. However, the calculated absolute magnitude with a Bulge distance hypothesis, $M_V\sim0$, is unusual. As the previous three RCBs star, it is more likely that EROS-CG-RCB-12 has an unusual thick circumstellar shell or is located behind the Bulge: with an estimated low temperature of $T_{eff}\sim4400$ K (which corresponds to an intrinsic V-I colour of $\sim1.9$), the maximum distance found is also of the order of $\sim14$ kpc.


After extinction correction, the Galactic DY Per-like RCB star EROS2-CG-RCB-2 has a  $V-I$ colour that corresponds to that of the Magellanic DY Per-like RCB stars (note also that the prototype DY Per star is very cool, $T_{\rm eff}\sim3500$, see \citet{1997PASP..109..969K}). However, it is $1-2$ magnitudes fainter than the DY Per-like RCB stars found in the SMC and LMC. This discrepancy may be due to EROS2-CG-RCB-2 being more distant, but correction for a (probable) additional extinction would make the star bluer and brighter.



Confirmation of the location of the stars could come from an equivalent width measurement of the interstellar lines of sodium (Na I, 5890.0, 5895.9 \AA{}) and/or potassium (K I, 7699 \AA{}) as there is a direct relation with reddening \citep*[see][]{1997A&A...318..269M}. The KI line should be more helpful due to the high extinction ($E(B-V)\geq0.4$) and also because NaI D absorption lines are usually observed during declines in RCB stars spectrum \citep{1996PASP..108..225C}. This study needs high resolution spectroscopy and has therefore not been performed with our data. We note that strong NaI D absorption lines are observed in our two highest resolution spectra of EROS2-CG-RCB-8 and EROS2-CG-RCB-10 as well as in most of the low resolution spectra in the left panel of Figure~\ref{spectrum_fig}.

We now comment on two issues related to the empirical clump giant method used to correct our magnitudes for extinction.  First, the calculated extinction is a regional average value, using clump giant stars located at less than
3 arcmin from each RCB.  If there is a localised low (high) extinction zone coinciding with the RCB, this method will over(under)-estimate the extinction. Second, most RCB stars have an intrinsic colour $V-I$ higher than that of the clump giant stars ($(V-I)^0_{Clump}\sim1.0$). The applied absorption coefficient $R_{B_E}=A_{B_E}/E(B_E-R_E)$ should depend of the spectrum slope, which is measured by the intrinsic star colour. Hence, use of $R_{B_E}$ derived from clump giants will not be exact for RCB stars. This effect may not be negligible, particularly with our wide EROS-2 bands (see \citet{2003A&A...404..423T} for an example of the variation of $R$ for Cepheids). We note that the small difference in slope in Figure~\ref{vvsvi}, between the Magellanic and Galactic Bulge RCBs, may be due to this effect: The 3 Bulge RCB stars with $V-I\sim1.0$ match well the position of the Magellanic RCBs in the colour magnitude diagram, whereas a small shift in luminosity is observed with the 6 redder Bulge RCBs, with $1.2< V-I <1.8$.



\subsection{Proper motion test}

We attempted to measure the proper motion for the 14 new RCBs using all available EROS-2 images, but no positive result was  obtained. The study was complicated by blending which causes correlations between flux and position. This is particularly evident for the RCB star EROS2-CG-RCB-14, which is listed in the NOMAD catalogue with a significant proper motion ($\mu_{\alpha} = 82\pm9$ and $\mu_{\delta}= 15.4\pm9$ mas/yr). We can explain this apparent movement by blending.

\subsection{Galactic latitude distribution}
\label{distrib}
We note that the density of RCB stars strongly increases towards the Galactic plane. We estimate a density in our fields of $\sim 0.6 \pm 0.2$ RCB per deg$^2$ for low Galactic latitude ($1\degr < \vert b \vert <3\degr$) and $\sim0.09 \pm 0.05$ RCB per deg$^2$ at higher latitude ($3\degr < \vert b \vert <5\degr$). For this calculation, only the 14 new EROS-2 RCB stars were used and all fields higher than $\vert b \vert >5\degr$ were not considered due to a dependence of our detection efficiency with Bulge RCBs absolute magnitude (see Sect.~\ref{detec_effi_dust})\footnote{We note also that a larger fraction of less frequently observed fields are located at $\vert b \vert >5\degr$ (see Fig.~\ref{cgfields})}.

For the latitude range $\vert b \vert <5\degr$, we applied the method of maximum likelihood to measure the scale height $h^{RCB}_{Bulge}$. We suppose a latitude distribution of the form exp($-b/h$). If we use the 14 new RCBs, we find $0.6<h^{RCB}_{Bulge}<2.75$ deg (95\% C.L.) with a most likely value of 1.0 deg : It corresponds at the distance of the Galactic Centre to a scale height of $137^{+239}_{-55}$ pc. But if we restrict our analysis to the 9 new RCB stars that are most likely to be located inside the Bulge, we found smaller $95\%$ C.L. limits: $0.45<h^{RCB}_{Bulge}<2.0$ deg with a most likely value of 0.8 deg, i.e. $109^{+137}_{-48}$ pc at the distance of the Galactic centre.

Such a small scale height is marginally consistent with the Bulge scale heights (270 to 360 pc) found by \citet{1995ApJ...445..716D} from the COBE/DIRBE data, using three different exponential triaxial Bulge models. The measured scale height is significantly less than that expected for a thick disk ($>1$ kpc) and is similar to that for young/intermediate age thin-disk populations.

The small statistic prevent us from having a clear view of the distribution of the new RCBs; however, their distances and scale height seem to indicate a disk-like distribution inside the Bulge. We note that such small scale heights have already been observed in the Galactic Bulge for OH/IR stars, which are oxygen-rich evolved stars \citep[see][]{1995A&A...299..689S,1999MNRAS.310..629S}. If confirmed, this result may help to constraint the age of the RCB stars, as a disk structure inside the Galactic Bulge could only be younger than the Galactic Bulge itself.

The observed scale height is lower than the one measured by \citet{1985ApJS...58..661I}, $\sim400$ pc, using known RCB stars outside the Galactic Bulge and assuming $M_{Bol} = -5$. This difference in scale height, between the inner part (Bulge) and the outer part (disk) of the Galaxy, has also been observed with OH/IR stars \citep{1999MNRAS.310..629S}.

No RCB stars have been found in the 29 spiral arms fields. This is not a significant result, as few RCB stars would be expected in these fields. Assuming a disk scale length of 2.5 kpc and the density found in our low Galactic latitude fields, we expected to find only $\sim2$ RCB stars in the spiral arms fields.


\subsection{Individual stars}

EROS2-CG-RCB-9, EROS2-CG-RCB-10 and EROS2-CG-RCB-11 all show multiple drops in brightness in their light curves, as well as carbon-rich spectra. They are undoubtedly members of the RCB class. None of these stars was previously catalogued as variable.

EROS2-CG-RCB-1 is the only new RCB stars that has not been confirmed spectroscopically. Its spectrum is almost totally featureless, although some weak CN bands are present; the star was very faint during the observation on JD-2454348. However its light curve has multiple declines of more than 8 magnitudes, making it a strong RCB candidate. After extinction correction, EROS2-CG-RCB-1 is the hottest RCB from the new sample. In fact, it's spectrum looks typical of a warm RCB star such as W Men \citep[see Fig.1][]{2001ApJ...554..298A} with a fair amount of interstellar extinction ($A_{B_E}=4.31$).

EROS2-CG-RCB-2 is a DY Per star candidate due to its slow decline and symmetric recovery, plus a location in the $J-H$ versus $H-K$ diagram (Fig.\ref{infra_red}) where all other DY Per stars are found. The spectrum confirms that the star is carbon-rich.

EROS2-CG-RCB-3 is the second hottest RCB star found in this analysis, after reddening correction. Its spectrum shows strong Ca II triplet lines as well as strong $C_2$ (Swan bands from 5500 to 6500 $\AA{}$) and CN features. The lack of measurements in the red band before the year 2000 is due to an electrical problem in CCD number 2.

EROS2-CG-RCB-4, named Terz V 2046 by \citet{1988A&AS...76..205T} was listed as a variable star due to its impressive luminosity change (close to 9 mag). Its R-band magnitude was measured at 9.2 in 1968, but fainter than 18 in 1981.

EROS2-CG-RCB-5 presents the quickest decline ($\sim$0.1 mag.day$^{-1}$) with a fast recovery 200 days later and the highest IR excess with $J-H \sim$ 2.74. Its spectrum shows strong Ca II triplet lines indicating that the star is hotter that the cool temperature estimated after a Bulge extinction correction.

EROS2-CG-RCB-6, named V1135 Sco in the GCVS\footnote{GCVS: General Catalogue of Variable Stars \citep{1971GCVS3.C......0K}}, was incorrectly classified as a Mira variable by \citet{1996A&AS..120..275A}. No indication of periodicity is seen in its light curve.

EROS2-CG-RCB-7 was named Terz V 1680 by \citet{1988A&AS...76..205T} with a variation in brightness of 5 magnitudes between 1980 and 1976. Multiple aperiodic drops in luminosity (more than 6 mag.) are visible in the EROS-2 light curve.

EROS2-CG-RCB-8 was named Terz V 2960 by \citet{1991A&AS...90..451T}. Its brightness varied by more than 3.3 magnitude between 1968 and 1987. No sudden drop was observed in the EROS-2 light curve, but with a total amplitude change of more than 7.6 magnitude in the $R_E$ band, a slow recovery of its maximum brightness (nearly 2.5 years) and a carbon type spectrum, EROS2-CG-RCB-8 is a strong RCB star candidate.

EROS2-CG-RCB-12 shows one characteristic decline around JD-2450900 along with many variations of large amplitude at maximum brightness. The total variation is slightly more than 3 magnitudes over the 6.7 years of observations. Its spectrum shows clear evidence of carbon molecules in its atmosphere, making this star a strong RCB candidate. Surprisingly, its spectrum shows no Ca II triplet lines indicating that this RCB star is the coolest of the classical new RCB star found.

EROS2-CG-RCB-13  has been monitored by both EROS-2 and the MACHO experiment (MACHO-176.19607.1138). The first 2 drops observed by EROS-2 are also present in the MACHO light curve, while EROS-2 shows another drop around the epoch JD-2451100.  Its spectrum indicates a carbon-rich atmosphere so it is also a strong RCB candidate.

EROS2-CG-RCB-14 has been catalogued in the NOMAD catalogue with a significant proper motion, which we ascribe to blending. This is the RCB star that shows the largest drop in luminosity in the new Galactic sample, with a variation of more than 8.3 magnitude in the $R_E$ band.

\section{Summary}
\label{summary}

Our search for new RCB stars in the EROS-2 Galactic photometry resulted in the discovery of 14 new RCB stars, one being a DY Per star. The total number of confirmed Galactic RCB stars is therefore now 51. No RCB was found in the 29 fields monitored in the spiral arms. Overall, we estimate that our detection of RCB stars should be $\sim$60\% complete for the majority of our survey.

MACHO-118.18666.100 has been misclassified as an RCB star. The new spectrum presented shows that it is an M giant star with light curve variations that resemble those of sequence-D variable stars.

If no interstellar reddening correction is applied, the new RCB sample would consist of the coolest RCB stars ever found. An extinction correction that assumes that the stars are located in the Bulge give intrinsic magnitudes and colours for 9 of the new RCBs that overlap those of the LMC and SMC. This indicates that they are indeed in the Galactic Bulge -- the first RCB stars identified in the Bulge.

Concerning the remaining classical RCBs, 4 have the lowest colour temperatures and faintest luminosities (EROS2-CG-RCB-5, -9, -12 and -14), assuming that the stars are located in the Bulge. However, these unusual properties may be explained either by an unusual thick circumstellar carbon shell or by a location beyond the Bulge at a maximum distance of $\sim14$ kpc. The supplementary distance would thus give rise to an extra reddening. From the study of the Ca II triplet strength, we suggest that EROS2-CG-RCB-5, -9, and -14 have typical RCB temperatures ($T_{eff}\sim5000$K) and that EROS2-CG-RCB-12 is the coolest of the classical RCBs found.


Finally, we note that the density of RCB stars seems to increase strongly as the galactic plane is approached. We measured a small scale height of $109^{+137}_{-48}$pc at the distance of the Galactic Centre using the 9 RCBs that are more likely to be located inside the Bulge. This small value is not consistent with measured Bulge scale height values and suggests a disk-like distribution inside the Bulge. Our derived distribution suggests that the RCB star density should be higher than one per deg$^2$ for $\vert b \vert <2\degr$ towards the Galactic Bulge. Hence, we suggest that the next generation of infrared variable star surveys of the Bulge (e.g. VISTA\footnote{Visible and Infrared Survey Telescope for Astronomy, URL: http://www.vista.ac.uk/}) should be able to discover many RCB stars near the Galactic Centre.

\begin{figure*}
\includegraphics[scale=0.45]{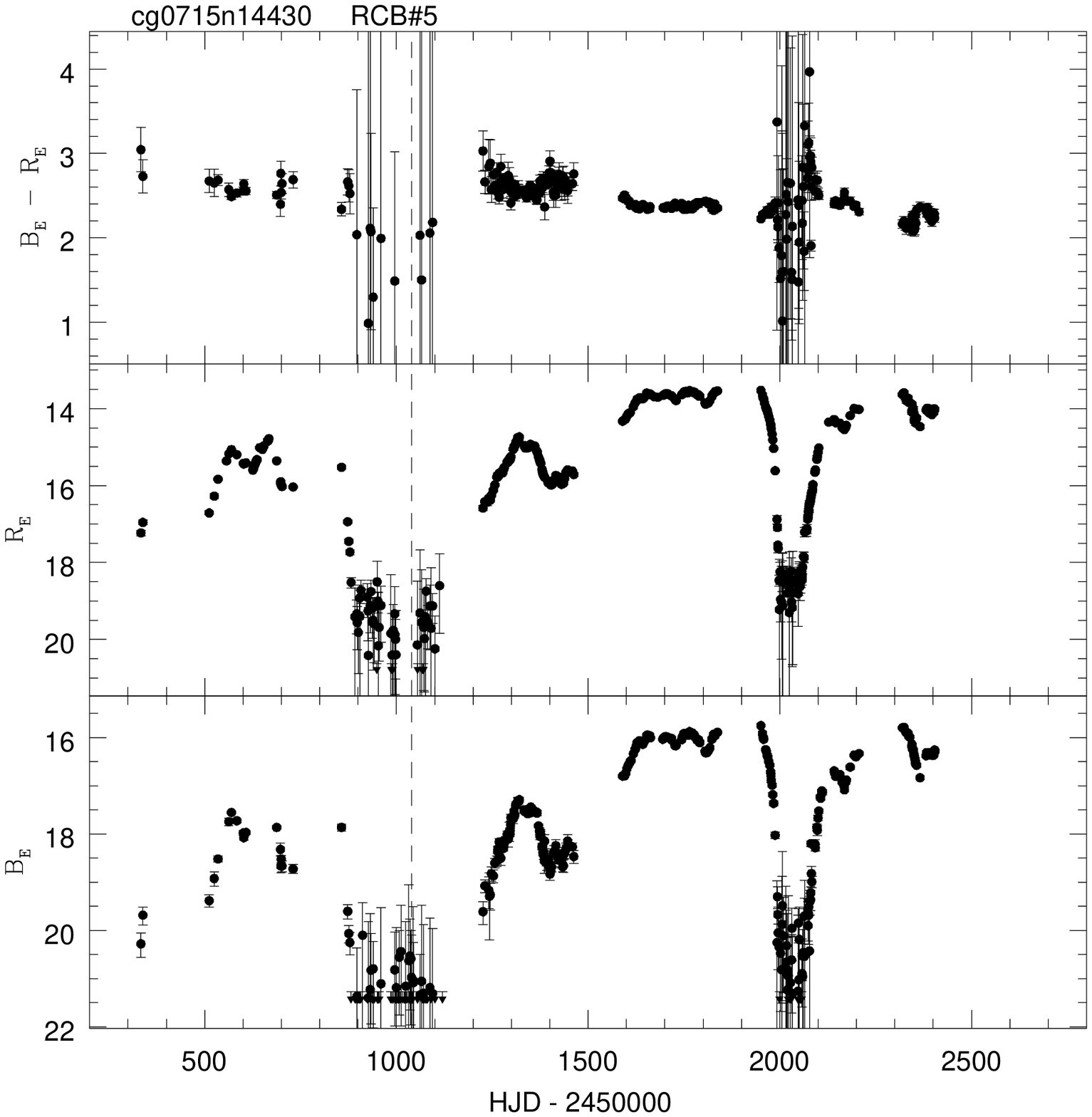}
\includegraphics[scale=0.45]{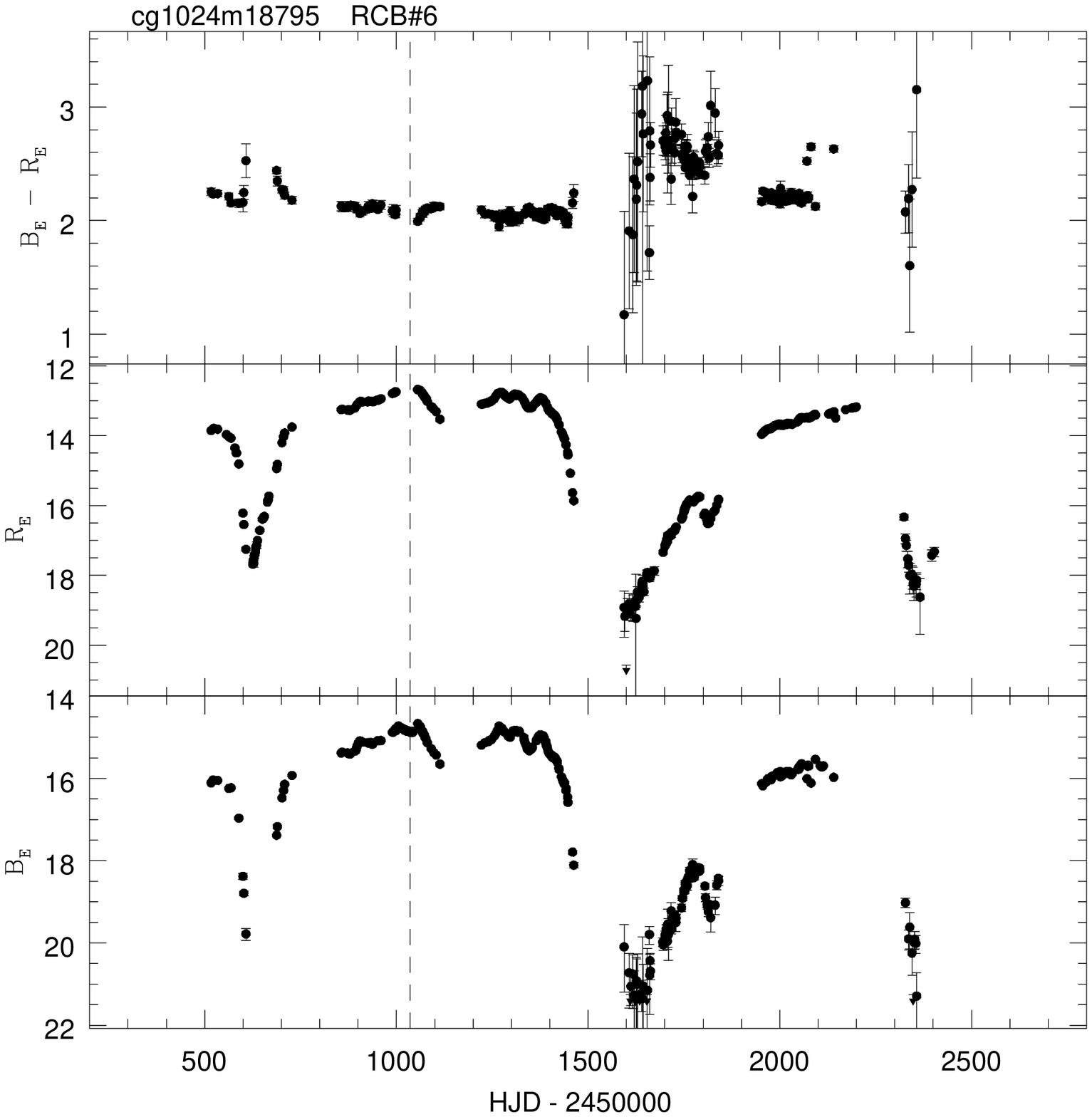}
\includegraphics[scale=0.45]{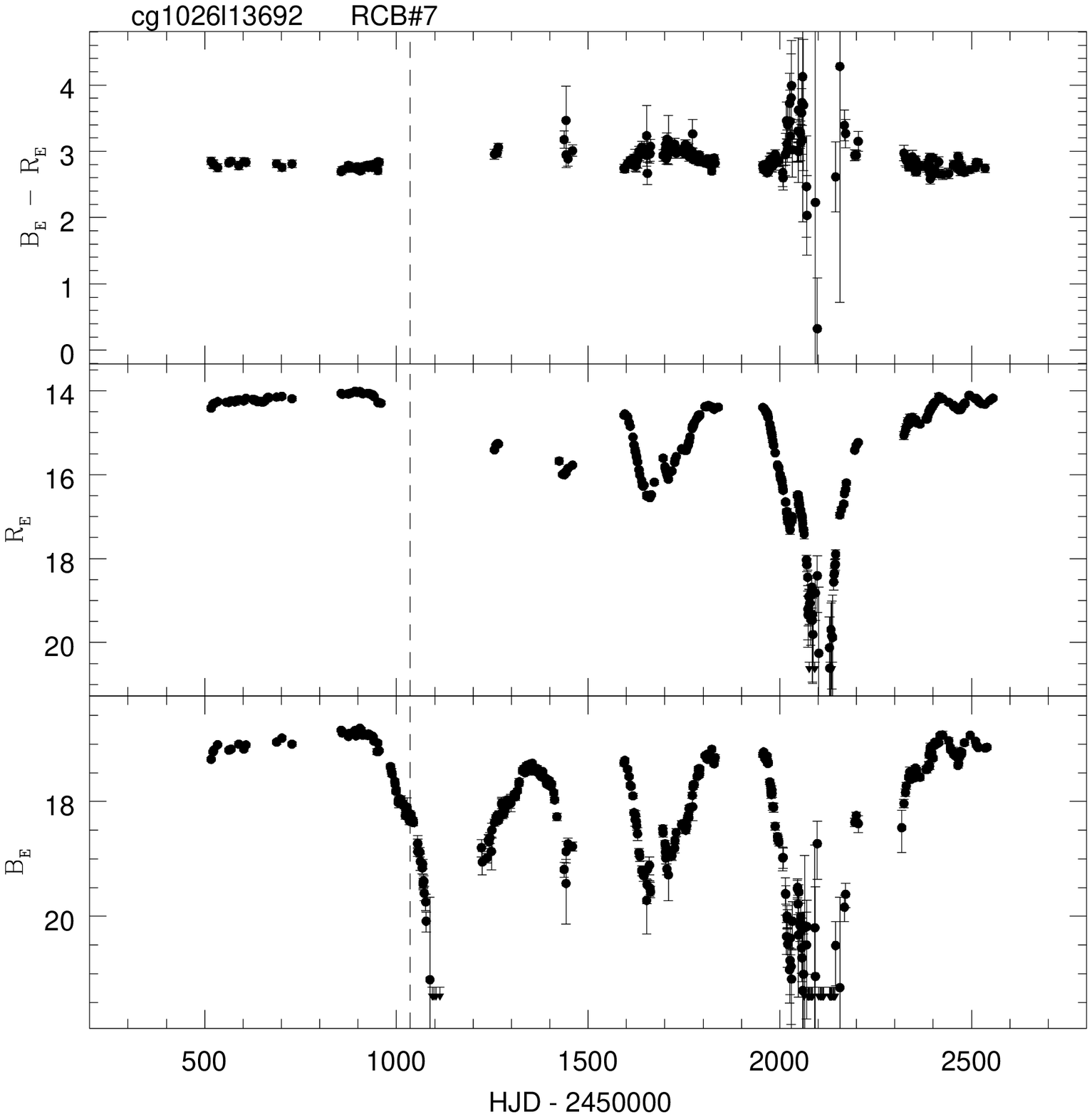}
\includegraphics[scale=0.45]{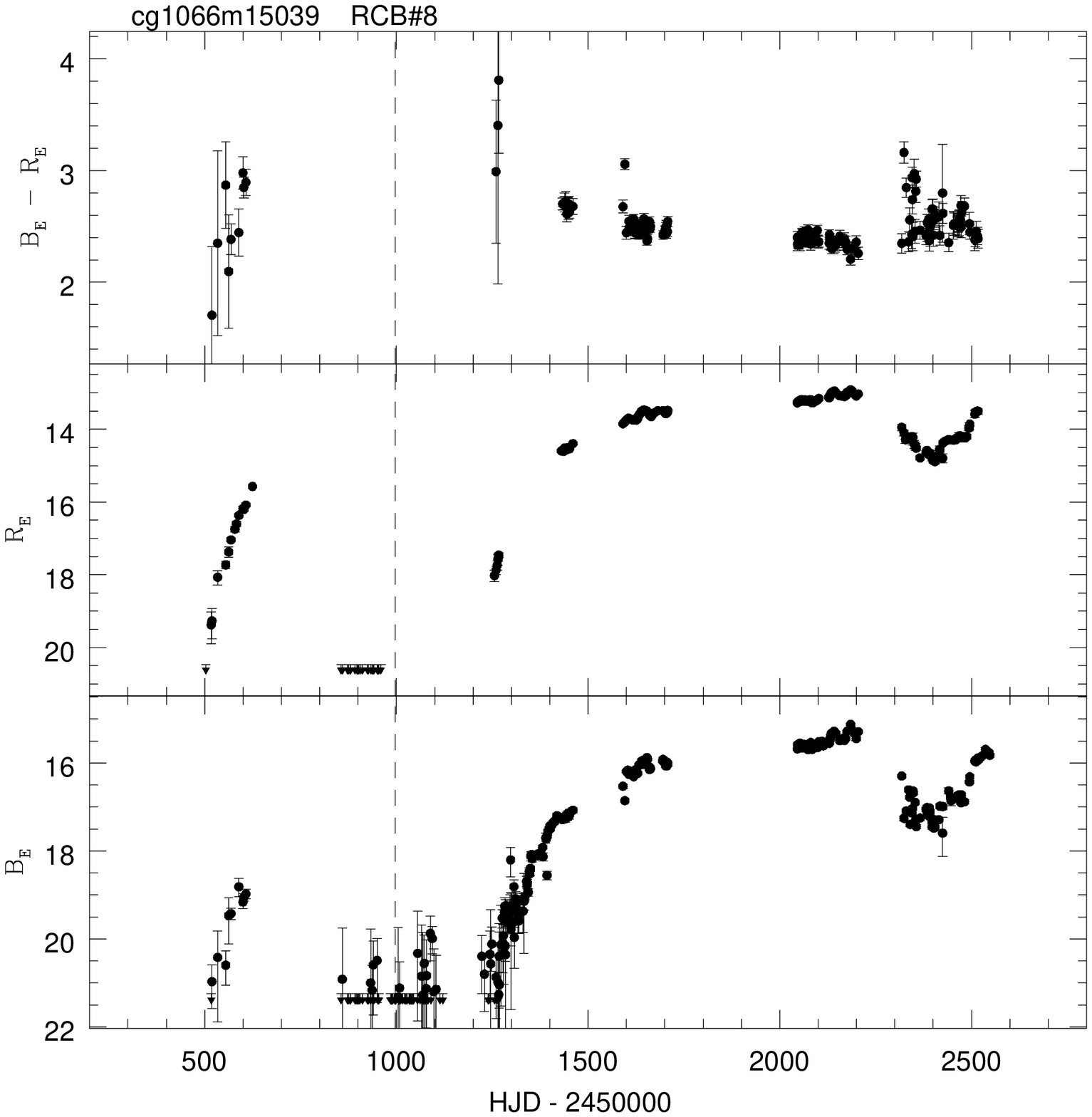}
\caption{Light curves of the new RCBs stars (continued).}
\end{figure*}

\begin{figure*}
\includegraphics[scale=0.45]{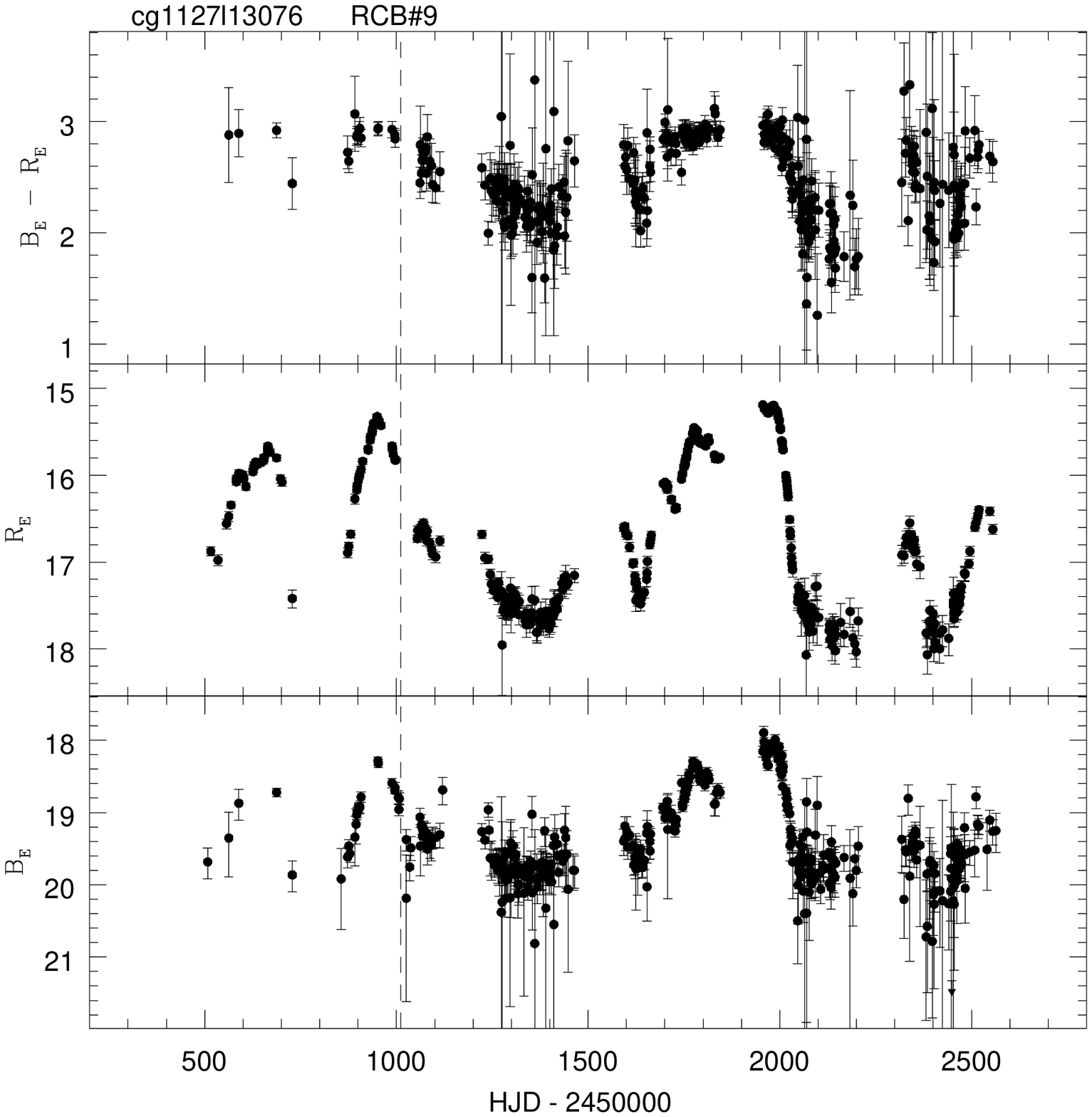}
\includegraphics[scale=0.45]{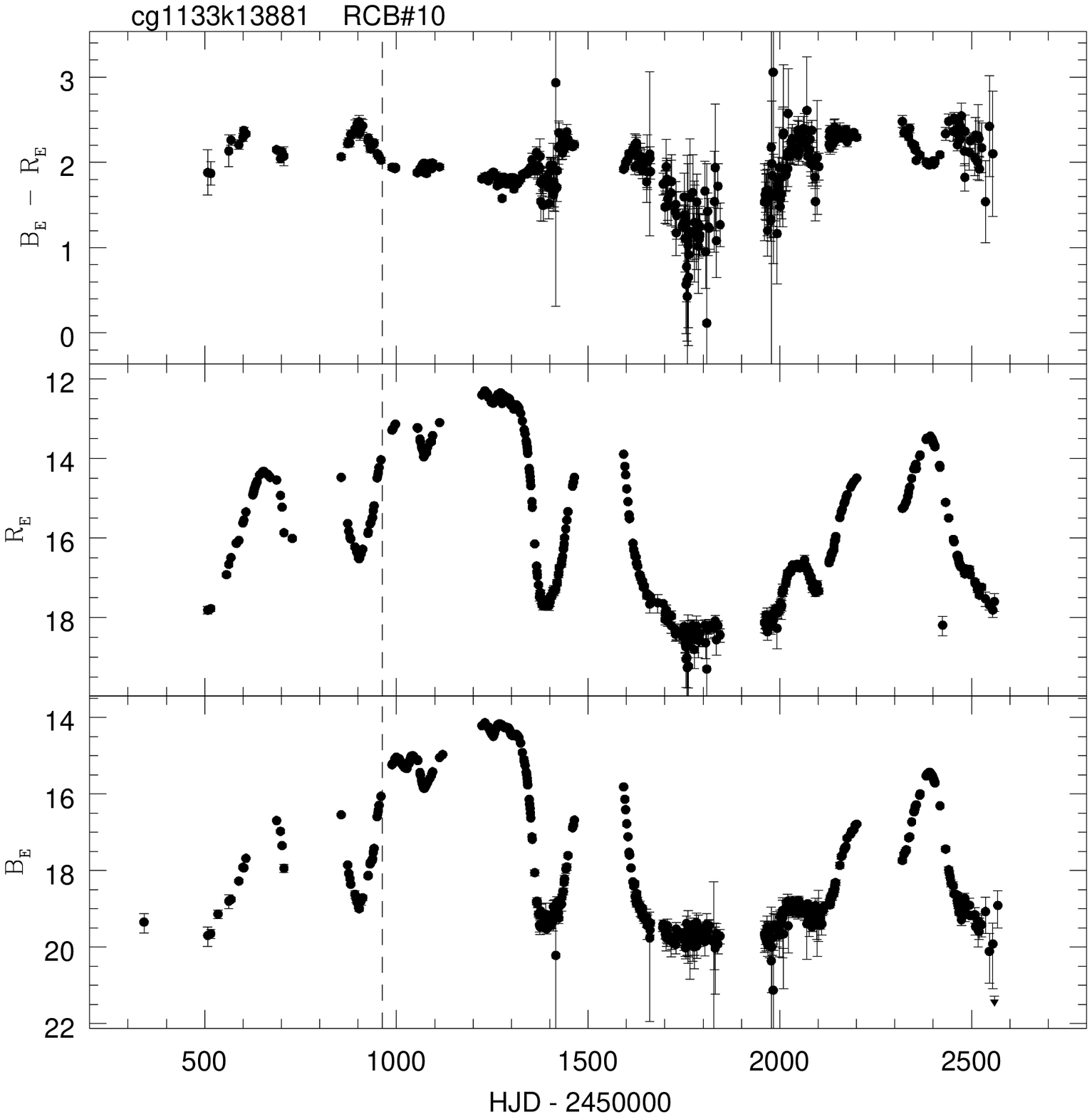}
\includegraphics[scale=0.45]{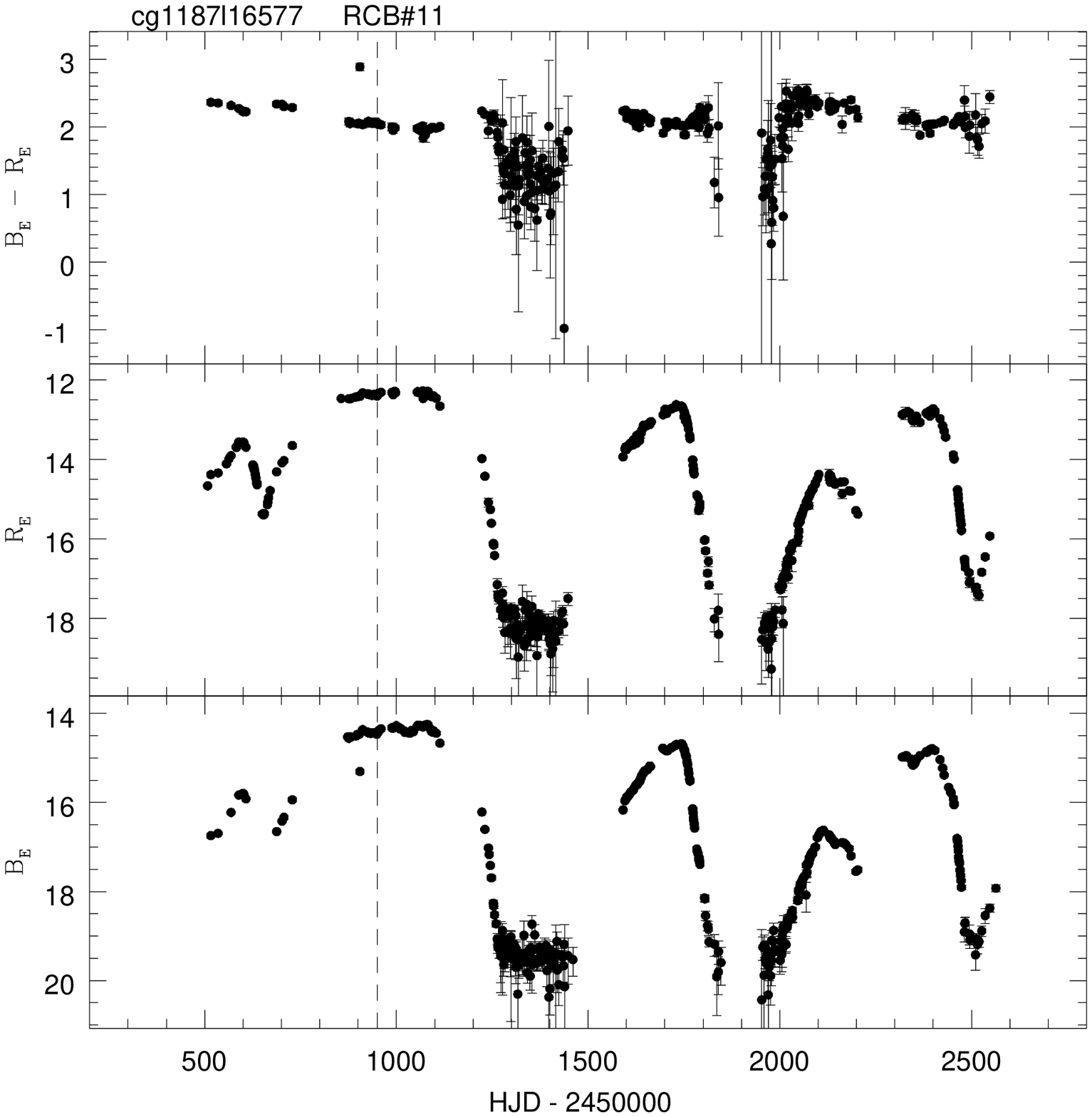}
\includegraphics[scale=0.45]{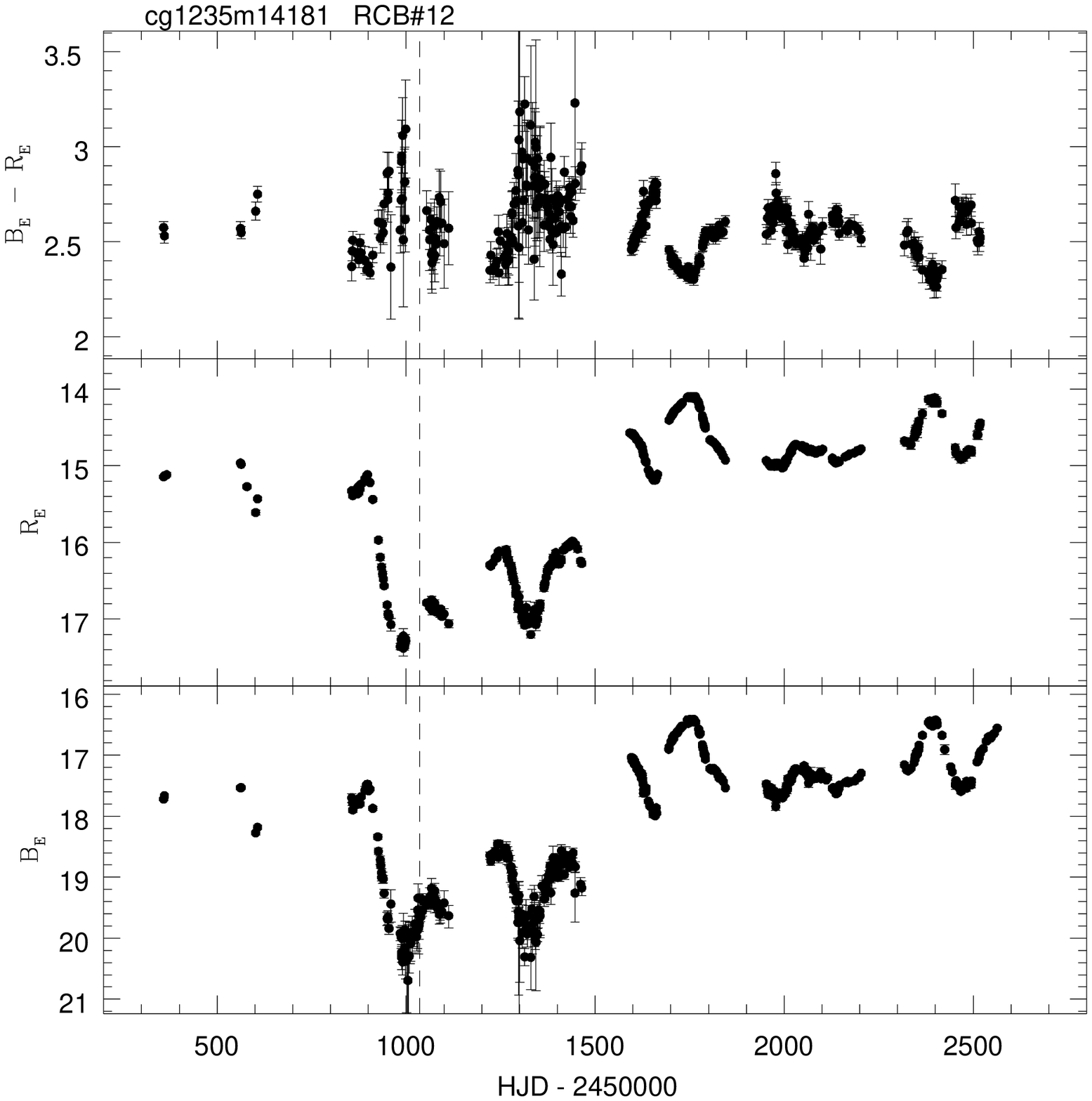}
\caption{Light curves of the new RCBs stars (continued).}
\end{figure*}

\begin{figure*}
\includegraphics[scale=0.45]{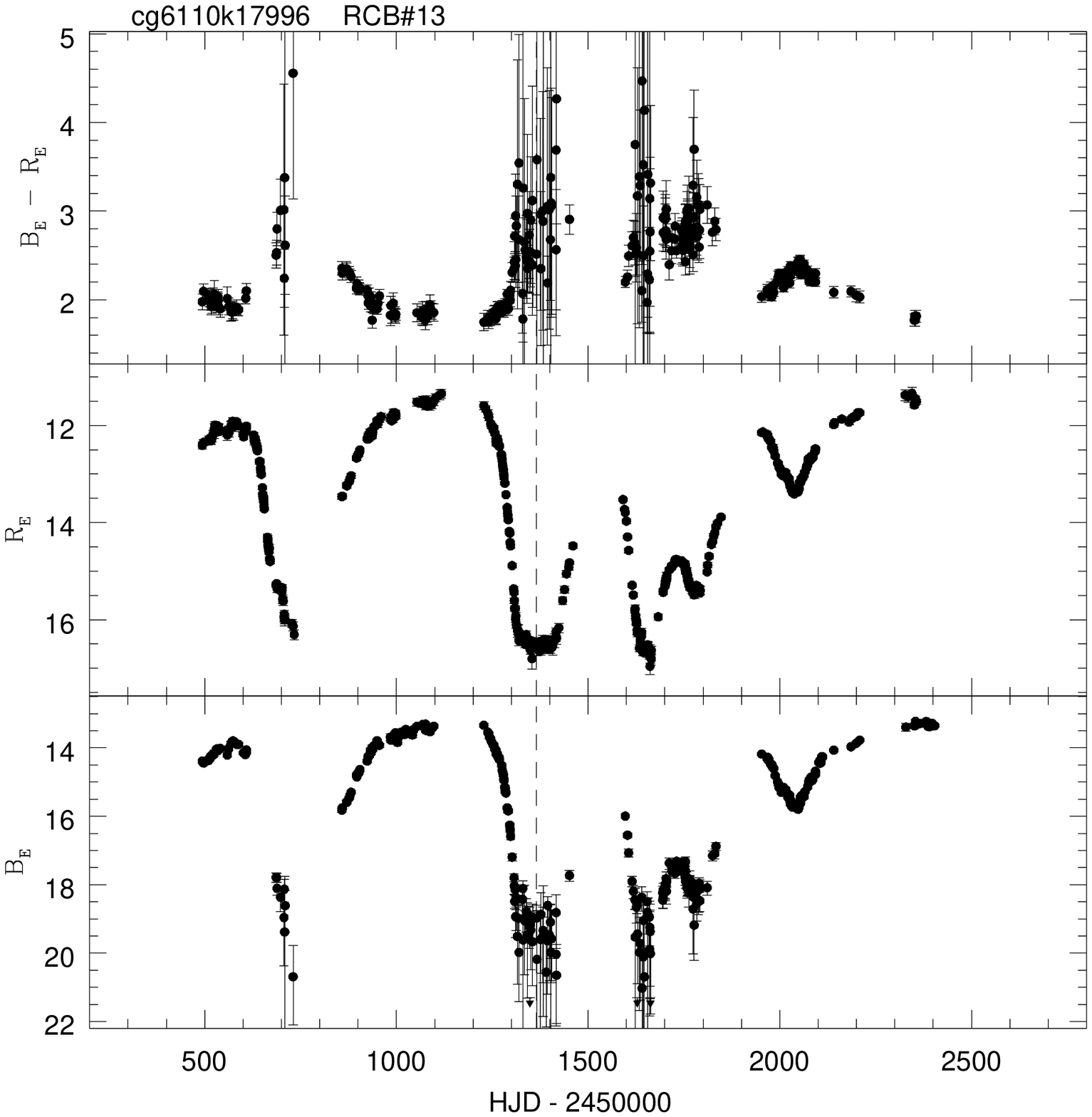}
\includegraphics[scale=0.45]{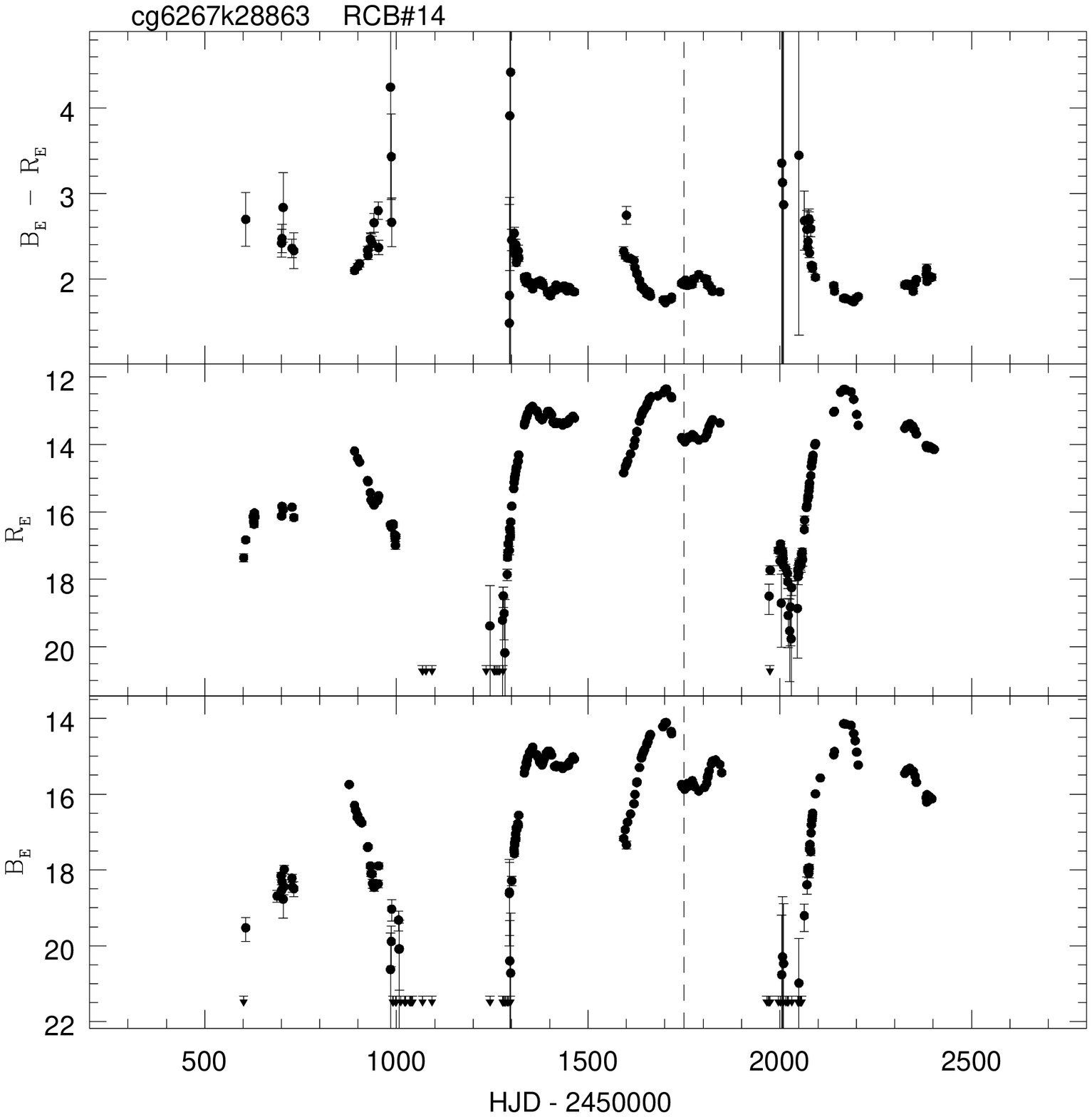}
\caption{Light curves of the new RCBs stars (continued).}
\label{lc_end}
\end{figure*}

\begin{figure*}
\centering
\includegraphics[scale=0.82]{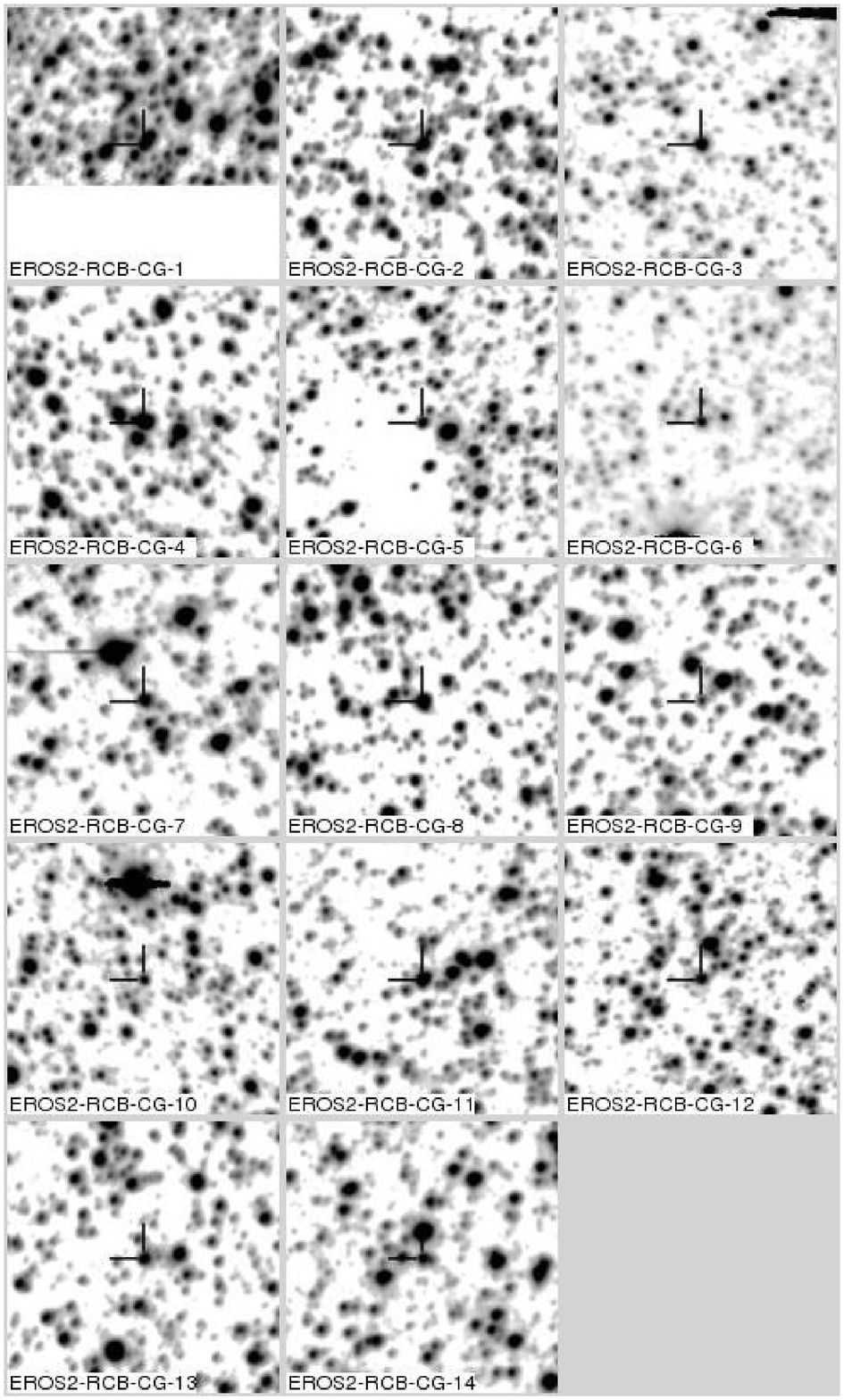}
\caption{Charts of the new Galactic RCB stars (2'x2'). North is up, East is to the left.}
\label{charts}
\end{figure*}

\begin{acknowledgements}
This paper is dedicated to the memory of Alain Milsztajn (1955-2007).

We would like to thank the referee, G.Clayton, for his comments that improved the paper and K.C.Freeman for useful discussions. This publication makes use of data products from the Two Micron All Sky Survey, which is a joint project of the University of Massachusetts and the Infrared Processing and Analysis Centre California Institute of Technology, funded by the National Aeronautics and Space Administration and the National Science Foundation. The Denis data have also been used. DENIS is the result of a joint effort involving human and financial contributions of several Institutes mostly located in Europe. It has been supported financially mainly by the French Institut National des Sciences de l'Univers, CNRS, and French Education Ministry, the European Southern Observatory, the State of Baden-Wuerttemberg, and the European Commission under networks of the SCIENCE and Human Capital and Mobility programs, the Landessternwarte, Heidelberg and Institut d'Astrophysique de Paris. JA acknowledges support from the Danish Natural Science Research Council.
\end{acknowledgements}

\bibliographystyle{aa}
\bibliography{RCB_CGBS}

\listofobjects

\end{document}